\documentclass[%
reprint,
showpacs,preprintnumbers,
 amsmath,amssymb,
 aps,
prx,
]{revtex4-1}
\usepackage{graphicx}
\usepackage{dcolumn}
\usepackage{bm}
\usepackage{hyperref}
\usepackage{amsmath}
\usepackage{amsfonts}
\usepackage{mathtools,bm}
\usepackage[amssymb]{SIunits}
\usepackage{color}
\usepackage{mathrsfs}
\usepackage{soul}
\usepackage{fancyhdr}
\usepackage{lineno}
\usepackage[yyyymmdd,hhmmss]{datetime}
\newcommand{\ud}[1]{{#1^{\dagger}}}

\begin{document}

\title{Polaritonic Rabi and Josephson Oscillations}
\author{A.~Rahmani}
\affiliation{Physics Department, Yazd University, P.O. Box 89195-741, Yazd, Iran}
\email{rahmani@stu.yazd.ac.ir}
\author{F.~P.~Laussy}
\affiliation{Russian Quantum Center, Novaya 100, 143025 Skolkovo, Moscow Region, Russia}
\affiliation{Condensed Matter Physics Center (IFIMAC), Universidad Aut\'{o}noma de Madrid, E-28049, Spain}
\email{fabrice.laussy@gmail.com}
\date{\today}

\begin{abstract}
  The dynamics of coupled condensates is a wide-encompassing problem
  with relevance to superconductors, BECs in traps, superfluids,
  etc. Here, we provide a unified picture of this fundamental problem
  that includes i) detuning of the free energies, ii) different
  self-interaction strengths and iii) finite lifetime of the modes. At
  such, this is particularly relevant for the dynamics of polaritons,
  both for their internal dynamics between their light and matter
  constituents, as well as for the more conventional dynamics of two
  spatially separated condensates. Polaritons are short-lived,
  interact only through their material fraction and are easily
  detuned. At such, they bring several variations to their atomic
  counterpart.  We show that the combination of these parameters
  results in important twists to the phenomenology of the Josephson
  effect, such as the behaviour of the relative phase (running or
  oscillating) or the occurence of self-trapping. We undertake a
  comprehensive stability analysis of the fixed points on a normalized
  Bloch sphere, that allows us to provide a generalized criterion to
  identify the Rabi and Josephson regimes in presence of detuning and
  decay.
\end{abstract}

\maketitle
\section{Introduction}

A superconductor can be described by an order parameter, that reduces
in the simplest formulation the dynamics of such a complex object to a
simple complex number~\cite{landau50a}. The question of what happens
with the phases of two superconductors put in contact through an
insulating barrier led Josephson to predict in 1962 with elementary
equations that a supercurrent should flow between them, driven by
their phase difference~\cite{josephson62a}.  The phenomenon was
quickly observed~\cite{anderson63a} and became emblematic of broken
symmetries and quantum effects at the macroscopic scale.  It was soon
speculated that a similar phenomenology should be observed with other
macroscopically degenerated quantum phases, such as superfluids or
Bose--Einstein condensates, even before the latter were experimentally
realized~\cite{javanainen86a}.  The role of the phase as the driving
agent of quantum fluids was brought to the fore by
Anderson~\cite{anderson66a} who identified ``phase slippage'' as a
source of dissipation~\cite{varoquaux15a}. Notably, in the case of
BECs, the first transposition of this physics was considering
non-interacting particles~\cite{javanainen86a} and the role of the
phase difference as a drive for the superflow was the focus of
attention. The question of the phase of macroscopically degenerate
quantum states remained anchored in the phenomenon but also took a
separate route of its own~\cite{leggett91a,castin97a,molmer97a}, that
is still actively investigated to this
day~\cite{anton14a,javanainen15a}.

The Josephson effect itself, on the other hand, was put on its
theoretical foothold by Leggett who defines it as the dynamics of~$N$
bosons ``restricted to occupy the same two-dimensional single particle
Hilbert space''~\cite{leggett01a}. Leggett introduced three regimes
for such systems depending on the relationship between tunelling and
interactions, namely the Rabi (non-interacting), Josephson
(weakly-interacting) and Fock (strongly-interacting)
regimes~\cite{leggett99a}.  ``Tunneling'' refers to linear coupling
between the condensates (quadratic in operators) while
``interactions'' refer to a nonlinear self-particle quartic term.  In
this sense, Josephson's physics is a limiting case of the
Bose--Hubbard model~\cite{veksler15a}, although the name retained a
strong bond with superconductors~\cite{barone_book82a}, possibly due
to the important applications it found as a quantum interference
device~\cite{jaklevic64a,kleiner04a} or merely for historical reasons
(the Josephson--Bardeen debate on the existence of the effect is one
highlight of scientific controversies~\cite{mcdonald01a}). To mark
this difference, one speaks of ``Bosonic Josephson junctions'' (BJJ)
for bosonic implementations of the Josephson dynamics~\cite{gati07a}.
This typically relates to condensates trapped in two wells, but due to
its fundamental and universal character as formulated by Leggett,
numerous platforms exhibit the effect. A pioneering report came from
superfluids~\cite{pereverzev97a}. For proper BECs, a so-called
``internal'' Josephson effect was deemed ``more promising'' with
alkali gases by involving different hyperfine Zeeman states rather
than a straightforward coupling between two spatially separated
condensates~\cite{leggett99a}. Eventually, the Josephson oscillation
was observed in a single junction of BEC~\cite{albiez05a}. In this
text, we consider another platform that can host Bose condensates:
microcavity polaritons~\cite{kavokin_book11a}. These systems having
demonstrated Bose--Einstein condensation~\cite{kasprzak06a} and
superfluid behaviour~\cite{amo09a}, are natural candidates to
implement the Josephson physics of coupled condensates---furthermore,
in strongly out-of-equilibrium open systems---and several theoretical
proposals have been made~\cite{sarchi08a,wouters08b,shelykh08a},
followed by experimental demonstrations, both in the linear
(Rabi)~\cite{lagoudakis10a} and nonlinear
(self-trapping)~\cite{abbarchi13a} regime. The polariton
implementation of Josephson effects is increasingly
investigated~\cite{aleiner12a,pavlovic13a,khripkov13a,racine14a,gavrilov14a,zhang14d,ma15a,rayanov15a,zhang15a}.
Recently, it was observed that polaritons are predisposed for
Josephson physics from the very nature of their light-matter
composition~\cite{voronova15b}, exhibiting the internal type of such
Josephson dynamics where the exchange is not between two spatially
separated condensates but between the two internal degrees of freedom
that make up the polariton, namely, its exciton and photon
components. This is an adequate picture, since condensates of
polaritons are also condensates of photons and
excitons~\cite{dominici14a}, with their Rabi coupling acting as the
tunneling. Interactions are then for the excitonic component only,
bringing a variation on the atomic counterpart in space, and detuning
of the free energies between the modes acts as the external potential,
so the analogy is essentially complete.  Bringing the framework of
Josephson dynamics to light-matter coupling sheds a new light on
polariton Rabi oscillations, in particular pointing out the phase
dynamics between the modes, which has been essentially ignored in the
description of such problems when considered at the level of coupled
oscillators~\cite{agarwal86a,carmichael89a,laussy09a} rather than
macroscopic wavefunctions. Since the latter are reduced to an order
parameter that does not need in most cases vary in space (but see
Refs.~\onlinecite{liew14a,colas16a}), both frameworks are tightly
related, and we explore such connections in the
following. Specifically, we focus on the general case where both
detuning and different on-site interactions are possible, also in
presence of decay and pumping, as befit light-matter interaction
problems in dissipative quantum optics. We show how this wider picture
blurs the line between Rabi and Josephson dynamics, or, rephrased more
positively, provides an elegant and natural physical picture that
brings the two regimes closer together.  We provide a general
criterion to take into account the new parameters and that should be
considered to claim the Josephson regime beyond the simple observation
of oscillations or of a running phase.

\section{Theory}

\begin{figure*}[th]
  \centering
  \includegraphics[width=.95\linewidth]{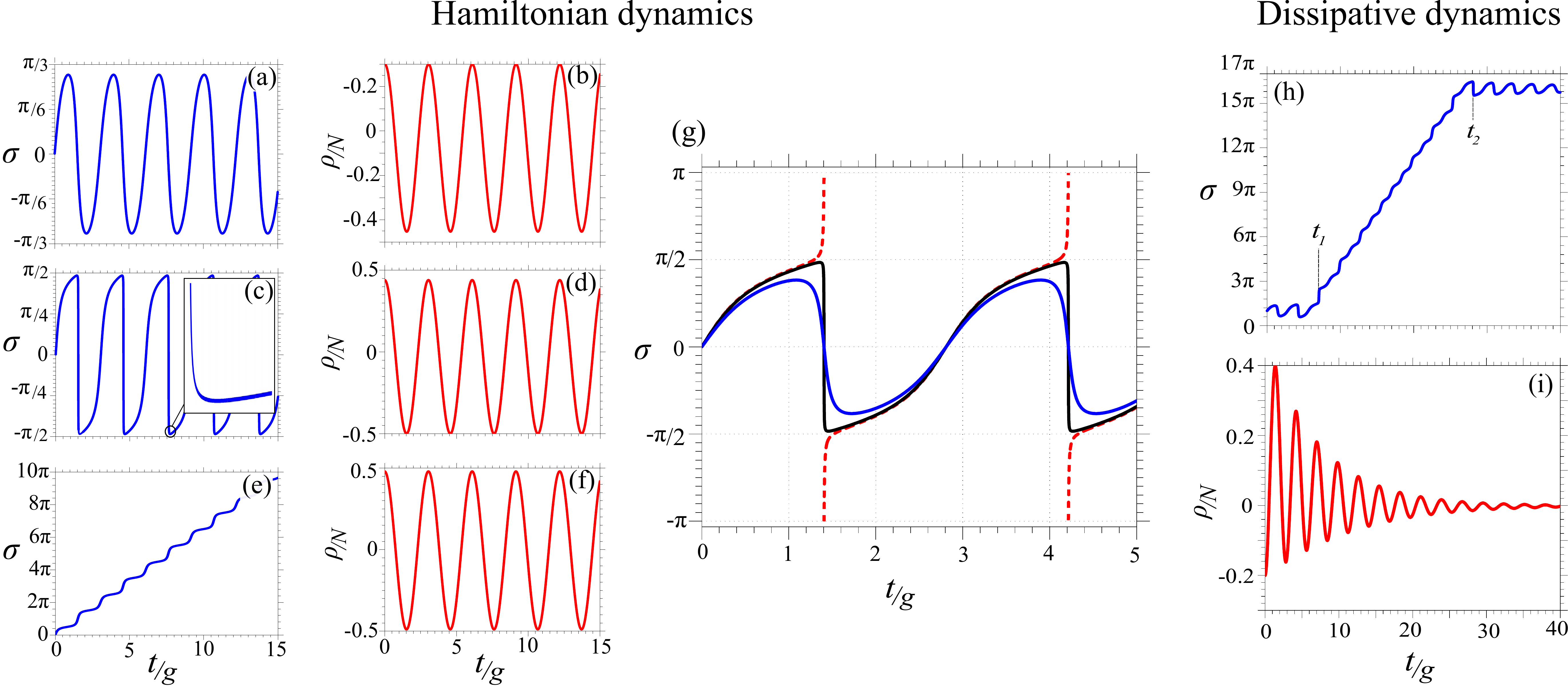}
  \caption{Dynamics of the relative phase~$\sigma$ and population
    imbalance~$\rho$ in a variety of scenarios of the pure Rabi
    regime: (a--b) oscillating-phase regime with sinusoidal
    cillations, with parameters $\rho_0=0.3N$, $\sigma_0=0$ and
    $\delta=-0.5$. (c--d) oscillating-phase regime with strongly
    anharmonic oscillations of the phase. A zoom in inset shows that
    the phase is continuous. The oscillations in the population remain
    sinusoidal. Parameters: $\rho_0=0.44N$, $\sigma_0=0$ and
    $\delta=-0.5$. (e--f) Running-phase regime (or discontinuous
    jumps if not unwrapped). Parameters: $\rho_0=0.48N$, $\sigma_0=0$
    and $\delta=-0.5$.  (g) Transition from the oscillating-phase
    (solid lines) to the running phase (dashed, not unwrapped)
    observed as a function of the population imbalance~$\rho_0$, from
    $\rho_0=0.25N$ (smooth blue curve) below threshold to
    $\rho_0=.305N$ (red--dashed line) above, passing by
    $\rho_0=0.299N$ (black, kinky line) very close to threshold (from
    below). Other parameters are: $\delta=-1$ and $\sigma_0=0$. (h)
    Relative phase and (i) population imbalance in the dissipative
    regime (with decay and without pumping nor interactions). There is
    a transition from the oscillating-phase to the running-phase
    regime at~$t_1$ and back at~$t_2$, with no notable feature in the
    population imbalance. Parameters: $\rho_0=-0.2N$, $\sigma_0=\pi$,
    $\delta=-1$, $\gamma_a=0.22g$ and $\gamma_b=.02g$, in which case
    $t_1\approx 6.9g$ and $t_2\approx28g$.}
  \label{fig:marjul7120100CEST2015123}
\end{figure*}

The dynamics of the Bosonic Josephson effect has been considered
extensively by Raghavan \emph{et al.}~\cite{raghavan99a} in a form
suitable for our discussion, including some considerations of
dissipation~\cite{marino99a} (see Ref.~\onlinecite{gati07a} for a
review).  We now briefly introduce the main points and notations. We
consider the coupling between two weakly-interacting Bose fields, $a$
(photons) and $b$ (excitons), with possibly different free
energies~$\epsilon_{a,b}$, ruled by the Hamiltonian:
\begin{subequations}
  \label{eq:marj282015611}
  \begin{align} 
    H&=H_0+V\,,\label{eq:marj2820156111}\\
    H_0&=\epsilon_a a^{\dagger}a+\epsilon_b b^{\dagger}b+g(a^{\dagger}b+b^{\dagger}a)\,,\label{eq:marj2820156112}\\
    V&=v_b(b^{\dagger}b^{\dagger}bb)+v_a(a^{\dagger}a^{\dagger}aa)\,.\label{eq:marj2820156113}
  \end{align}
\end{subequations}

Of course our results and conclusions apply to other systems than the
internal Josephson dynamics of light-matter coupling, as long as they
are well described by Eqs.~(\ref{eq:marj282015611}), but we will keep
this terminology for convenience. In a Josephson workframe, $a$
and~$b$ are ground state annihilation operators and the
averages~$\langle a\rangle$ and~$\langle b\rangle$ are order
parameters ($c$-numbers) for the two condensates (we will
note~$n_a\equiv\langle\ud{a}a\rangle$
and~$n_b\equiv\langle\ud{b}b\rangle$ the populations of each
mode). The dynamics can be described in terms of i) the population
imbalance $\rho\equiv(\langle a^{\dagger}a\rangle-\langle
b^{\dagger}b\rangle)/2 =(n_a-n_b)/2$ between the two modes and ii)
their relative phase~$\sigma=\arg{\langle a^{\dagger}b\rangle}$. Note
that the relative phase is, strictly speaking, $S\equiv\arg(\langle
a\rangle-\langle b\rangle)$ while we define it here as~$\sigma$, the
argument of a first-order cross-correlation. This is done for greater
generality as it allows us to describe all types of quantum states for
coupled harmonic oscillators, including mixed states, as will be
discussed in section~\ref{sec:marjul7120100CEST201521}. For coherent
states (describing ideal condensates), $S=\sigma$, and our convention
thus causes no loss of generality. Note that such mean-field
approximations that provide the pillars for the physics at play have
been relaxed in recent years and exact (numerical) solutions are now
available~\cite{sakmann09a,chuchem10a} that, interestingly, depart
considerably from the established picture, in particular regarding the
role of the phase. In our mean-field approximation, where $\langle
a^{\dagger}bb^{\dagger}b\rangle\approx\langle
a^{\dagger}b\rangle\langle b^{\dagger}b\rangle$ and $\langle
a^{\dagger}ba^{\dagger}a\rangle\approx\langle
a^{\dagger}b\rangle\langle a^{\dagger}a\rangle$ (this assumes that the
states remain coherent states), the two observables are ruled by the
following equations of motion~\cite{raghavan99a,voronova15b}:
\begin{subequations}
  \label{eq:marj282015615}
  \begin{eqnarray}
    \partial_t(\rho/N)&=F_1(\rho,\sigma)\equiv&-\sqrt{1-4(\rho/N)^2}\sin(\sigma)\,,\\
    \partial_t\sigma&=F_2(\rho,\sigma)\equiv&\Delta E-2(\rho/N)\Lambda+{}\nonumber\\
    &&{}+\frac{4\rho/N}{\sqrt{1-4(\rho/N)^2}}\cos(\sigma)\,,
  \end{eqnarray}
\end{subequations}
where we introduce the notation~$F_{1,2}$ for future convenience.
$N\equiv\langle a^{\dagger}a\rangle+\langle b^{\dagger}b\rangle$ is
the total number of particles, $\delta=(\epsilon_a-\epsilon_b)/g$ is the
bare modes detuning and we highlight two particular parameters of
importance to describe the dynamics, an effective detuning~$\Delta E$
and an effective blueshift~$\Lambda$:
\begin{subequations}
  \begin{align}
    \Delta E&\equiv -\delta+N(v_b-v_a)/g\,,  \label{eq:domfeb28110505CET2016}\\
    \Lambda&\equiv(v_a+v_b)N/g\,. \label{eq:marmar8144817CET2016}
  \end{align}
\end{subequations}
Equations~(\ref{eq:marj282015615}) are the so-called BJJ
equations~\cite{gati07a} that describe the dynamics of coupled
BECs. They differ in several aspects from Superconducting Josephson
Junction equations but also bear enough resemblances to lead to
similar physics. Voronova \emph{et al.}~\cite{voronova15b} recently
reported a peculiar phase dynamics of BJJ when including detuning,
even in the linear regime: the phase oscillations are strongly
anharmonic and possibly even get in a regime of phase-jumping (or
freely running phase if unwrapped). This is noteworthy as reminiscent
of the Josephson dynamics, i.e., driven by interactions. Without
interactions, oscillations in populations remain harmonic for all
detunings, indeed with some renormalization of frequencies and nonzero
imbalance, as can be expected from the conventional Rabi coupling
picture out of resonance.  Particular cases of this time dynamics
for~$\rho$ and~$\sigma$ are shown in
Fig.~\ref{fig:marjul7120100CEST2015123}. In panel~(a) and~(c),
$\sigma$ oscillates in time, while in panel~(e), $\sigma$ is
running. Note how in panel~(c), close to the frontier between the
two-cases, the phase is highly deformed from harmonic oscillations. An
inset shows that the oscillation is continuous even though it becomes
very sharp. This is also seen in
Fig.~\ref{fig:marjul7120100CEST2015123}(g) for three regimes around
the transition, showing how the phase abruptly changes from steep
oscillations of~$\pi$ amplitudes (solid lines) to~$2\pi$ phase jumps
(dashed line).  In panel~(e), we have unfolded the phase for clarity.
In all cases, particles transfer harmonically between the two states
as oscillations in~$\rho$ show. For such pure Hamiltonian dynamics,
initial conditions as well as detuning determine the possible regimes
of relative phase. There follows a rich phase diagram that can be
characterized analytically~\cite{voronova15b} in the linear or weakly
interacting regime. Here it must be stressed again that the same
phenomenology that is usually attributed to Josephson dynamics is
observed without interactions, that is, in the pure Rabi regime. This
calls to reconsider what is meant, precisely, by Josephson and Rabi
dynamics. We clarify this point below.

In the quantum-optical mindset, dissipation is an essential part of
the dynamics~\cite{agarwal84a,carmichael89a}. This is also an
ingredient that is important to describe short-lived polaritons. To do
so, the formalism is upgraded from an Hamiltonian to a Liouvillian
description, leading to a master equation for the density
matrix~$\varrho$:~\cite{carmichael_book02a,delvalle_book10a}:
\begin{eqnarray}\label{eq:gh12324556}
  \partial_t\varrho=i[\varrho,H]+&&\sum_{c=a,b}\frac{\gamma_c}{2}(2c\varrho c^{\dagger}-c^{\dagger}c\varrho-\varrho c^{\dagger}c)+\nonumber\\&&\sum_{c=a,b}\frac{p_c}{2}(2c\varrho c^{\dagger}-c^{\dagger}c\varrho-\varrho c^{\dagger}c)\,,
\end{eqnarray}
where~$\gamma_c$~and~$p_c$~are decay and incoherent pumping rates for
the states~$c=a,b$. Then, Eqs.~(\ref{eq:marj282015615}) in this
dissipative regime and for coherent states become:
\begin{subequations}
  \label{eq:marjul7120733CEST201145666}
  \begin{align}
    \partial_t(\rho/N)&=-\sqrt{1-4(\rho/N)^2}\sin(\sigma)+\frac12\Gamma_--2(\rho/N)^2\Gamma_-\,,\\
        \partial_t\sigma&=\Delta E-2(\rho/N)\Lambda+\frac{4\rho/N}{\sqrt{1-4(\rho/N)^2}}\cos(\sigma)\,,
  \end{align}
\end{subequations}
where (with~$p_a=p_b=0$ in the case with only decay):
\begin{equation}
  \label{eq:domfeb28122646CET2016}
  \Gamma_\pm\equiv\frac{1}{2}[(p_a-\gamma_a)\pm(p_b-\gamma_b)]\,.
\end{equation}

One example of the dissipative Rabi dynamics is shown in
Fig.~\ref{fig:marjul7120100CEST2015123}(h--i). Remarkably, one observes
a switching in time between the oscillating and running phase regimes.
This switching is caused by the decay and depends on the initial
condition as well as detuning.  The switching happens if the
occupation in one state becomes exactly zero, with all the population
residing in the other state, although this is not a necessary
condition. When this occurs, the phase of the emptied state becomes
ill-defined and so does the relative-phase. Such a change of regime
can appear two times, as shown in the figure, a single time or none.
In any case, the system always ends up in the regime of oscillating
phase, except at resonance where the running mode can last forever.
There is therefore only one switching when the initial condition and
detuning are such that the dynamics starts in the running phase mode.
These are mere statements of the facts. We will explain the reason for
this peculiar behaviour in the following and it will become clear that
such an apparently rich phenomenology is in fact trivial and bears no
connection to Rabi and Josephson dynamics.

Equations~(\ref{eq:marj282015615})
and~(\ref{eq:marjul7120733CEST201145666}) are the traditional form for
the (bosonic) Josephson dynamics, coupling population imbalance and
relative phase in a way that supports the notion that one drives the
other. This form conceals, however, the more fundamental structure
that underpins the relationship between the key variables: population
imbalance indeed, but the full complex
correlator~$\langle\ud{a}b\rangle$ rather than merely its phase (once
the connection is understood, however, one can indeed limit to the
phase). The dynamics is thus put in full view geometrically in the
$(\rho, \langle\ud{a}b\rangle)$ space. Since $\langle\ud{a}b\rangle$
is complex, the space is three-dimensional. The trajectories can be
obtained from the full set of equations:
\begin{subequations}
  \label{eq:marjul7120100CEST20159876}
  \begin{align}
    \partial_t\langle a^{\dagger}b\rangle=(i(\delta)+\Gamma_+)\langle a^{\dagger}b\rangle-i\langle a^{\dagger}a\rangle+i\langle b^{\dagger}b\rangle\nonumber\\-2iv_b\langle a^{\dagger}bb^{\dagger}b\rangle+2iv_a\langle a^{\dagger}ba^{\dagger}a\rangle\,,\\
    \partial_t\langle a^{\dagger}a\rangle=-i\langle a^{\dagger}b\rangle+i\langle b^{\dagger}a\rangle+(p_a-\gamma_a)\langle a^{\dagger}a\rangle+p_a\,,\\
    \partial_t\langle b^{\dagger}b\rangle=i\langle a^{\dagger}b\rangle-i\langle b^{\dagger}a\rangle+(p_b-\gamma_b)\langle b^{\dagger}b\rangle+p_b\,.
  \end{align}
\end{subequations}
Diagonalizing these equations, we get one key result for the dynamics:
\begin{equation}
  \label{eq:marjul7120100CEST201522}
  |\langle\ud{a_\theta}b_\theta\rangle|^2+\rho_\theta^2=({N(t)}/{2})^2+\mathcal{P}(t)\,,
\end{equation}
where:
\begin{equation}
  \mathcal{P}(t)=-\exp(2\Gamma_+t)\int_0^t\exp(-2\Gamma_+t')(p_an_b+p_bn_a)dt'\,.
\end{equation}
This result holds even in the interacting case (with $v_a\neq0$ and/or
$v_b\neq0$), but since the Hamiltonian then needs be diagonalized at
all times, this is just a formal way to rewrite the equation. In other
cases, the geometric nature of the dynamics is captured. Here, the
main, albeit obvious, argument is the introduction of the generic
equation for~$\theta$, the \emph{mixing angle} between exciton and
photons, describing a change of basis:
\begin{subequations}
  \label{eq:marjul7120100CEST20159899}
  \begin{align}
    a_{\theta}&=\cos(\theta)a+\sin(\theta)b\,,\\
    b_{\theta}&=-\sin(\theta)a+\cos(\theta)b\,,
  \end{align}
\end{subequations}
\begin{figure*}[t]
  \centering
  \includegraphics[width=.9\linewidth]{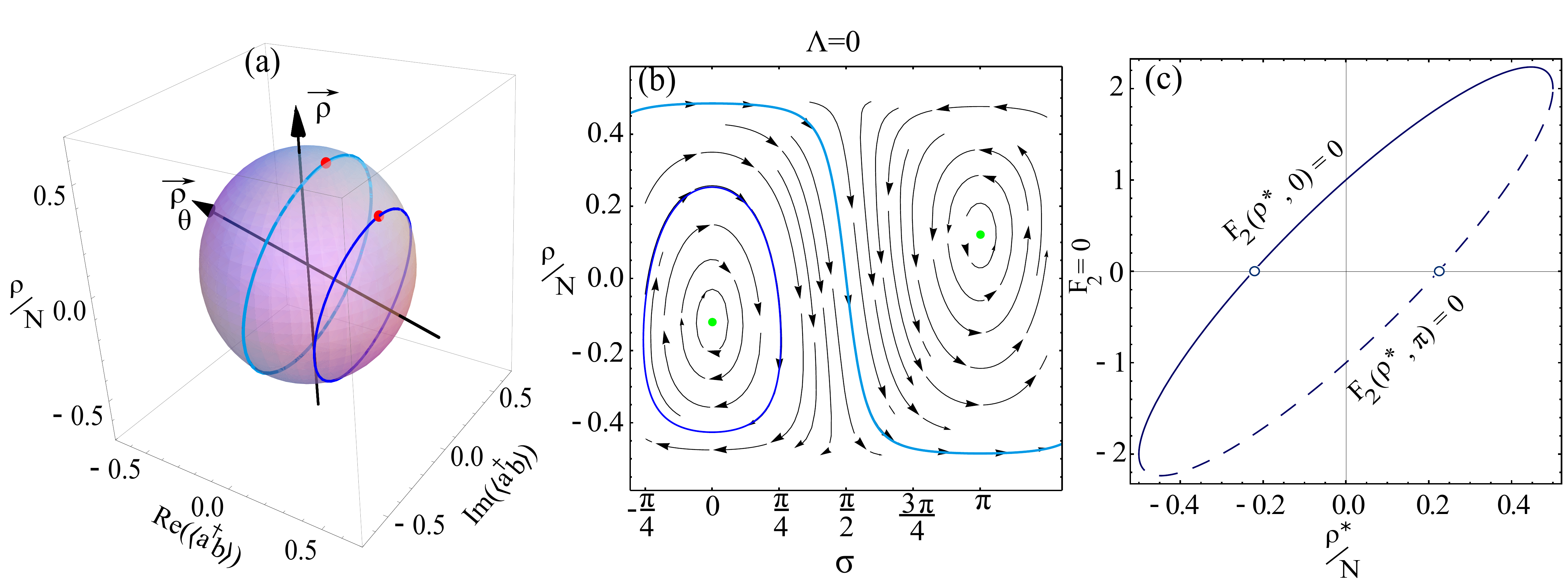}
  \caption{(a) The dynamics of two coupled condensates is clearly
    understood on a Bloch sphere. The polariton basis defines an
    axis~$\vec\rho_\theta$ around which the pure Rabi dynamics evolves
    as a simple circle, whose distance from the center is determined
    by the quantum state, with polaritons at the poles and the
    full-amplitude Rabi oscillations between the dressed states as the
    equator. At resonance, $\delta=0$, $\vec\rho$ and
    $\vec\rho_\theta$ are orthogonal. parameters are the same as
    Fig~\ref{fig:marjul7120100CEST2015123}. (b) Projection of the
    dynamics of the two cases in panel~(a) on the $(\rho,\sigma)$
    space, superimposed on the streamlines of the dynamical
    system. There are two centers, displayed as green dots, located
    at~$\sigma=0$ and $\sigma=\pi$. (c) The fixed points are solutions
    of $F_2(\rho^*,\sigma^*)=0$, that is, the intersection of the
    curve with the~$x$ axis, indicated by open circles. The solid line
    corresponds to $\sigma=0$ and the dashed one to $\sigma=\pi$. The
    detuning was taken as $\Delta E=0.5$.}
  \label{fig:marjul7120100CEST20151158}
\end{figure*}
where~$\cos(\theta)=\sqrt{1/2+\delta/2\sqrt{4+\delta^2}}$. These
operators diagonalize the Hamiltonian~(\ref{eq:marj2820156112}) and
lead to the simple solution,
Eq.~(\ref{eq:marjul7120100CEST201522}). In the $(\rho,
\langle\ud{a}b\rangle)$ space, the trajectory is therefore simply that
of ``circles on a sphere''. In non-Hamiltonian cases, the radius
changes in time but solutions remain equally simple if kept on a
normalized sphere. This sphere is a counterpart of the Bloch sphere,
that describes the dynamics of a two-level system. Here the two levels
are the exciton and photon amplitudes~\cite{dominici14a} which yield
the parametrization for the Bloch vector~$\mathbf{v}$:
\begin{equation}
  \label{eq:jueene21113822CET2016}
  \mathbf{v}=\frac{2}{N}\left(
    \frac{\langle a^{\dagger}b+b^{\dagger}a\rangle}{2i},
    \frac{\langle a^{\dagger}b-b^{\dagger}a\rangle}{2i}
    \frac{\langle a^{\dagger}a-b^{\dagger}b\rangle}{2}
  \right)\,.
\end{equation}
The typical representation in a $(\rho, \sigma)$ plane~\cite{gati07a}
produces instead complex patterns, even in the linear case of simple
circular motion, as a result of the transformation involved by
projecting from a sphere. In recent years, it has however become more common
to represent the Josephson dynamics on its appropriate
geometry~\cite{zibold10a,chuchem10a,arXiv_voronova16a}.

\section{Hamiltonian regime}
\label{sec:marjul7120406CEST2015}

\subsection{Dynamics}

We first revisit the usual Hamiltonian case with no dissipation, i.e.,
when $p_a=p_b=\gamma_a=\gamma_b=0$. The pure Rabi regime, when $v_a$
and~$v_b$ are zero, admits analytical solution for~$\rho$ and the
phase~\cite{voronova15b}. Namely, for the population imbalance~$\rho$:
\begin{multline}
  \label{eq:marjul7120100CEST20152}
    \rho(t)=\big\lbrace\delta\big(2\cos(\sigma(0))+\rho(0)\delta)+{}\\{}+(4\rho(0)-\delta P(0)\cos(\sigma(0))\big)\cos(Rt)+{}\\{}+RP(0)\sin(\sigma(0))\sin(Rt)\big\rbrace/R^2\,,
\end{multline}
where we introduced $P\equiv\sqrt{N^2-4\rho^2}$ and
$R\equiv\sqrt{4+\delta^2}$. As for the relative phase $\sigma(t)$ (it
can also be obtained from the real and imaginary parts
of~$n_{ab}=\langle a^{\dagger}b\rangle$):
\begin{equation}
  \label{eq:marjul7120100CEST20150}
  \sigma=-\sin^{-1}(\partial_t\rho/P)\,.
\end{equation}
This is an exact, albeit obscure, description of the dynamics that is
put in full view geometrically on the Bloch sphere.  Indeed, the
comparison between the particular case
Eqs.~(\ref{eq:marjul7120100CEST20152}--\ref{eq:marjul7120100CEST20150})
with the general solution~Eq.~(\ref{eq:marjul7120100CEST201522}) shows
the great simplification brought by the geometric representation. The
nonlinear case has no closed-form solution to the best of our
knowledge although as a two-dimensional dynamical system, its solution
are readily obtained numerically (we provide separately an applet to
compute the trajectories in both the sphere and projected on the
phase-space~\cite{rahmani16Wolfram}). From
Eqs.~(\ref{eq:marjul7120100CEST20152}--\ref{eq:marjul7120100CEST20150}),
for instance, one can derive the conditions for oscillating or running
phase, by considering whether $\sigma(t)$ is bounded in time, in which
case the function is oscillating. This is achieved by finding zeros
for its derivatives, leading to the following equation for the
frontier between the two regimes of phase dynamics as a function of
detuning and initial conditions:
\begin{eqnarray}
  \label{eq:marjul7120100CEST2015119865}
  \rho(0)=N\frac{4\cos^2(\sigma(0))-\delta^2}{2(4\cos^2(\sigma(0))+\delta^2)}\,.
\end{eqnarray} 
If $\rho(0)$ is less than the rhs, then the phase is oscillating,
otherwise it is running.  This perplexing result is easily understood
on the Bloch sphere, as shown in
Fig.~\ref{fig:marjul7120100CEST20151158}(a), where the Rabi dynamics
reduces to a simple circle. This circle is concisely and fully
described by its normal axis~$\vec\rho_\theta$ and its
distance~$\rho_\theta$ from the equator in this basis. The latter is
given, on the normalized sphere, by:
\begin{equation}
  \label{eq:marene12230820CET2016}
  \rho_\theta=\left(\rho\delta+\frac{2\mathrm{Re}\langle\ud{a}b\rangle}{R}\right)\frac{1}{RN}\,.
\end{equation}
This is a familiar result in quantum-optical terms. In the proper
basis---of dressed states---the dynamics is that of the free
propagation (circular motion) of uncoupled states. This is determined
by the~$\vec\rho_\theta$ axis, around which the dressed states evolve
freely (harmonically) at a distance from the equator that is
determined by their state (their content of lower and upper
polaritons), leading to a linearly increasing phase. We can now
explain that what determines the dynamics of the phase (oscillating or
running) is simply whether the trajectory on the sphere encircles or
not the South--North $\rho$ axis defined by the laboratory observables
(i.e., of the bare states). In the proper basis of dressed states, the
phase is always running. Bare states on the other hand are the
familiar physical objects of the system in which terms it is
convenient to think. In our case, they are the exciton and photon
modes, and are furthermore those typically observed experimentally
(only the photons in most experiments).  This laboratory basis is, in
the case of optimal strong-coupling, orthogonal to the dressed state
basis, with $\vec\rho_\theta\perp\vec\rho$, and the circular motion is
observed as a sinusoidal oscillation (a circle observed sidewise) in
the general case, or even a saw-tooth function when the quantum state
maximizes the amplitudes of oscillations by satisfying $\rho/N=\pm1/2$
(for instance starting with all polaritons in one mode at~$t=0$).
This is shown in Fig.~\ref{fig:marjul7120100CEST2015123}(a,b)
and~(e,f) that correspond to the case of
Fig.~\ref{fig:marjul7120100CEST20151158}(a), namely, as initial
conditions: a 50-50 (light blue) superposition of $\theta$-eigenstates
and another ratio (in blue), leading to a smaller circle, both normal
to the~$\vec\rho_\theta$ axis. As observed in the exciton-photon
basis, their~$\rho$ and~$\sigma$ dynamics is distorted. There is no
such distortion for the population imbalance, since the circular
motion from any circle on the sphere projected on any normal axis
still results in a sine function. However the relative phase is
defined by that of the vector that joins the center of the sphere and
the circle itself. If the circle lies outside the axis, the phase can
remain always unequivocally defined in a~$2\pi$ interval, leading to
oscillations as the trajectory reaches an apex on the circle and turns
back. This is the situation of the blue circle in
Fig.~\ref{fig:marjul7120100CEST20151158}(a). In the other case where
the circle goes round the axis, there is no turning point and the
phase increases forever. This is the situation of the light blue
circle in Fig.~\ref{fig:marjul7120100CEST20151158}(a).  It is clear,
then, that the dynamics of the phase has no deep meaning of driving a
flow of particles. Instead, it pertains to a choice of basis. The
oscillating phase regime corresponds to a case where the basis of
observables is too far apart from that which is natural for the system
and the tilt between them is so large that the phase is distorted into
a qualitative different behaviour of oscillations instead of a linear
drift. In contrast, the running phase regime is that where the system
is described by observables close to the dressed states of the system.

The rationale of Leggett in distinguishing between a Rabi and
Josephson regime was to set apart the cases where tunneling~$g$ (in
our case, Rabi coupling) dominates from that where nonlinearity~$v$
dominates.  At resonance and for equal interaction on both sites, the
exact criterion is to compare $2vN/g$ to unity, ($N$ the total number
of particles, $v$ the self-interaction and $g$ the tunneling
strength). This is indeed correct but, even in absence of dissipation,
is restricted to resonance and equal nonlinearities. Here we provide
the general result to set apart the Rabi and Josephson regimes in
presence of detuning, which is required in general since detuning may
fake a Josephson-looking dynamics even in non-interacting systems.

\subsection{Classification of fixed points}

As we are dealing with a dynamical system, the standard procedure to
classify the possible trajectories is a stability analysis around the
fixed points. In the BJJ, the fixed points~$\rho^*$ and~$\sigma^*$ are
by definition the solutions $F_{i}(\rho^*,\sigma^*)=0$ for~$i=1,2$
(cf.~Eqs.~(\ref{eq:marj282015615})). There are two possible solutions
for the phase, $\sigma^{\ast}=0$ and~$\sigma^{\ast}=\pi$
(modulo~$2\pi$, so that~$\sigma^{\ast}=-\pi$ is also a solution in a
closed~$2\pi$ interval). Solving for the other variable, we exhaust
the possible fixed points. Their stability is determined by the
eigenvalues~$\lambda_i$ of the Jacobian Matrix given by:
\begin{equation}
  \label{eq:miémar16191926CET2016}
  J=\left( \begin{array}{*2{c}}‎\partial_{\rho}F_1&\partial_{\sigma}F_1\\
      \partial_{\rho}F_2&\partial_{\sigma}F_2 \end{array}\right) _{(\rho^{\ast},\sigma^{\ast})}\,,
\end{equation}
and the type can be mapped on a diagram with
axes~$\Delta\equiv\lambda_1\lambda_2$ and
$\tau\equiv\lambda_1+\lambda_2$~\cite{strogatz_book94a}, as shown in
Fig.~\ref{fig:marjul7120100CEST83140}.

\subsubsection{Non-interacting case}

First, in the non-interacting case, the system admits simple
closed-form solutions:
\begin{subequations}
  \begin{align}
    \label{eq:juefeb11174734CET2016}
    \sigma^*=0\quad\text{and}\quad\rho^*=\frac{N}2\frac{\delta}{\sqrt{4+\delta^2}}\,,\\
    \sigma^*=\pi\quad\text{and}\quad\rho^*=-\frac{N}2\frac{\delta}{\sqrt{4+\delta^2}}\,.
  \end{align}
\end{subequations}

As is clear on physical grounds, detuning can produce a state with a
large population imbalance, which can bear resemblance to macroscopic
quantum self-trapping even in absence of interaction. Using the
definition of Raghavan \emph{et al.}~\cite{raghavan99a} that the system
is macroscopically self-trapped when its total energy balances the
coupling strength, we can find a critical detuning that satisfies this
condition in absence of interactions, namely:
\begin{equation}
\label{eq:juefeb11174734CET3132016}
\delta_s=\frac{1-\sqrt{1-4(\rho(0)/N)^2}\cos(\sigma(0))}{\rho(0)/N}\,.
\end{equation}   

Examples of dynamics on the Bloch sphere in this noninteracting case
are shown in Fig~\ref{fig:marjul7120100CEST20151158}(a) and on the
projected $(\rho,\sigma)$ space in
Fig.~\ref{fig:marjul7120100CEST20151158}(b), with the two fixed points
at~$\sigma=0$ and $\sigma=\pi$ marked by (green) points. The two
orbits show the running and oscillatory phases surrounding these fixed
points without being attracted nor repelled by them. In the
terminology of dynamical systems, this corresponds to fixed points
that are neutrally stable.  Geometrically, the fixed points are the
intersections with the $x$-axis of the curves shown in
Fig.~\ref{fig:marjul7120100CEST20151158}(c) (zeros of~$F_2$).

\begin{figure}[tb]
  \centering
  \includegraphics[width=\linewidth]{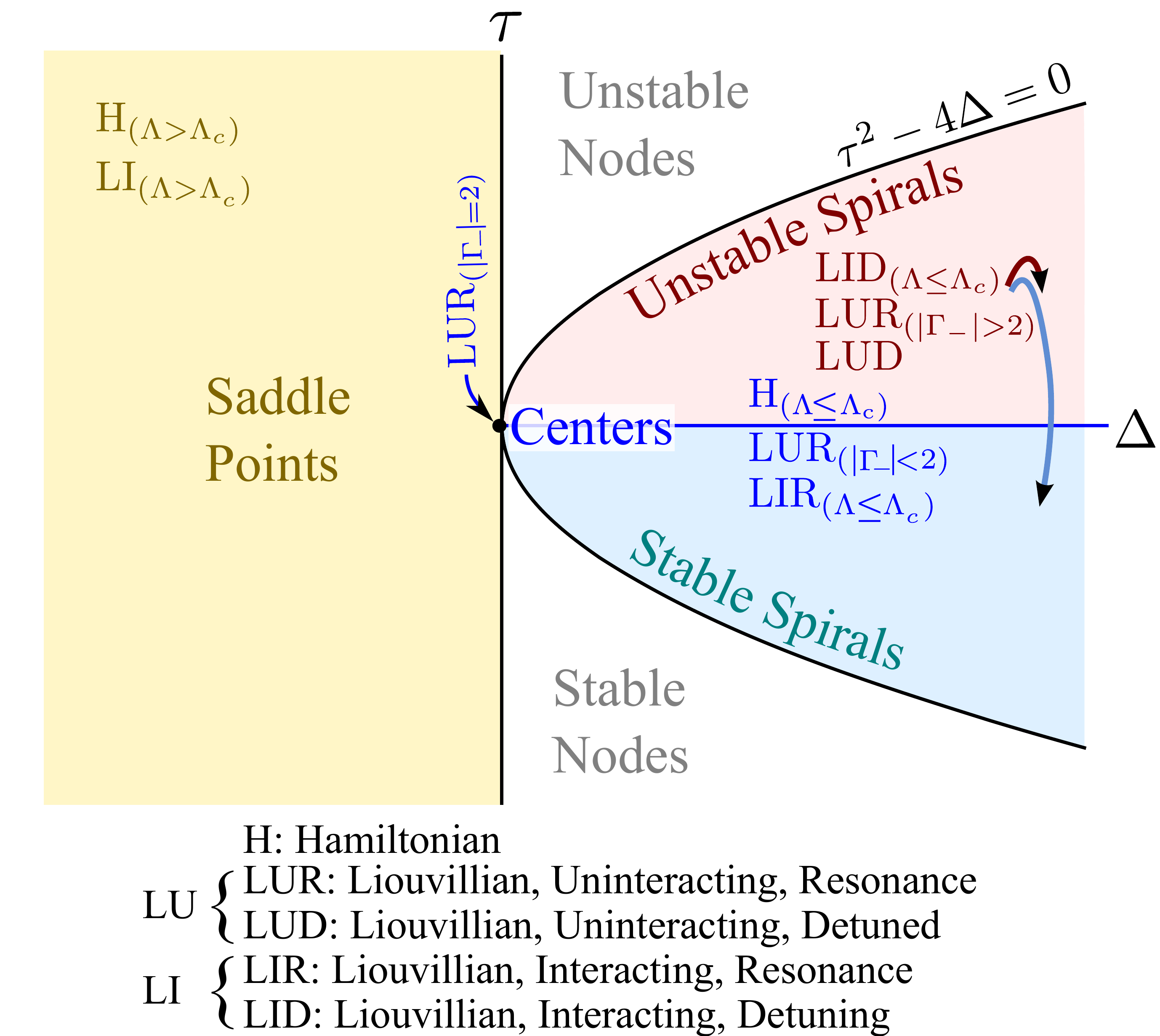}
  \caption{Classification of the fixed points of the dissipative
    Bosonic Josephson Junction. The axes are the functions
    $\Delta\equiv\lambda_1\lambda_2$
    and~$\tau\equiv\lambda_1+\lambda_2$ of the Jacobian's eigenvalues,
    cf.~Eq.~(\ref{eq:miémar16191926CET2016}). Our terminology is
    spelled out at the bottom of the figure, with (at most) three
    letters to label each case: first letter is either H (Hamiltonian)
    or L (Liouvillian) for the cases without or with decay,
    respectively. Second letter is U (uninteracting) or I
    (interacting) for the cases without or with self-interactions,
    respectively. Third letter is R (resonance) or D (detuned) for the
    cases $\delta=0$ or $\delta\neq0$, respectively. Further criteria
    are specified as subscripts. For instance,
    $\mathrm{LI}_{\Lambda>\Lambda_c}$ are dissipative systems with
    interactions with both zero and nonzero detuning such that
    $\Lambda>\Lambda_c$, in which case these systems have fixed points
    with saddle instability. The fact that $\Lambda>\Lambda_c$ is
    equivalent to the existence of a saddle fixed point allows to use
    the latter as a necessary and sufficient criterion for the
    Josephson regime.  The Nodes area, separated from the Spirals
    by~$\tau^2-4\Delta=0$, are not accesible. The purple and blue
    arrows for the points $\mathrm{LUD}_{(|\Gamma_-|<2)}$ and
    $\mathrm{LID}_{(\Lambda\leq\Lambda_c)}$ mean that these points
    belong to both the Unstable and Stable Spirals regions. All the
    cases shown here are for~$\Delta E=0$.}
  \label{fig:marjul7120100CEST83140}
\end{figure}

For the stability that follows from
Eq.~(\ref{eq:miémar16191926CET2016}), we find
$\lambda_1=i\sqrt{4+\Delta E^2}$ and $\lambda_2=-i\sqrt{4+\Delta
  E^2}$, implying that $\Delta>0$ and~$\tau=0$. As a consequence, the
two fixed points in the Rabi regime are centers, i.e., they are stable
and every near-enough trajectory is
closed~\cite{strogatz_book94a}. These are the
$\mathrm{H}_{(\Lambda\leq\Lambda_c)}$ points in
Fig.~\ref{fig:marjul7120100CEST83140} with~$\Lambda=0$.

\begin{figure*}[!htbp]
  \centering
  \includegraphics[width=.85\linewidth]{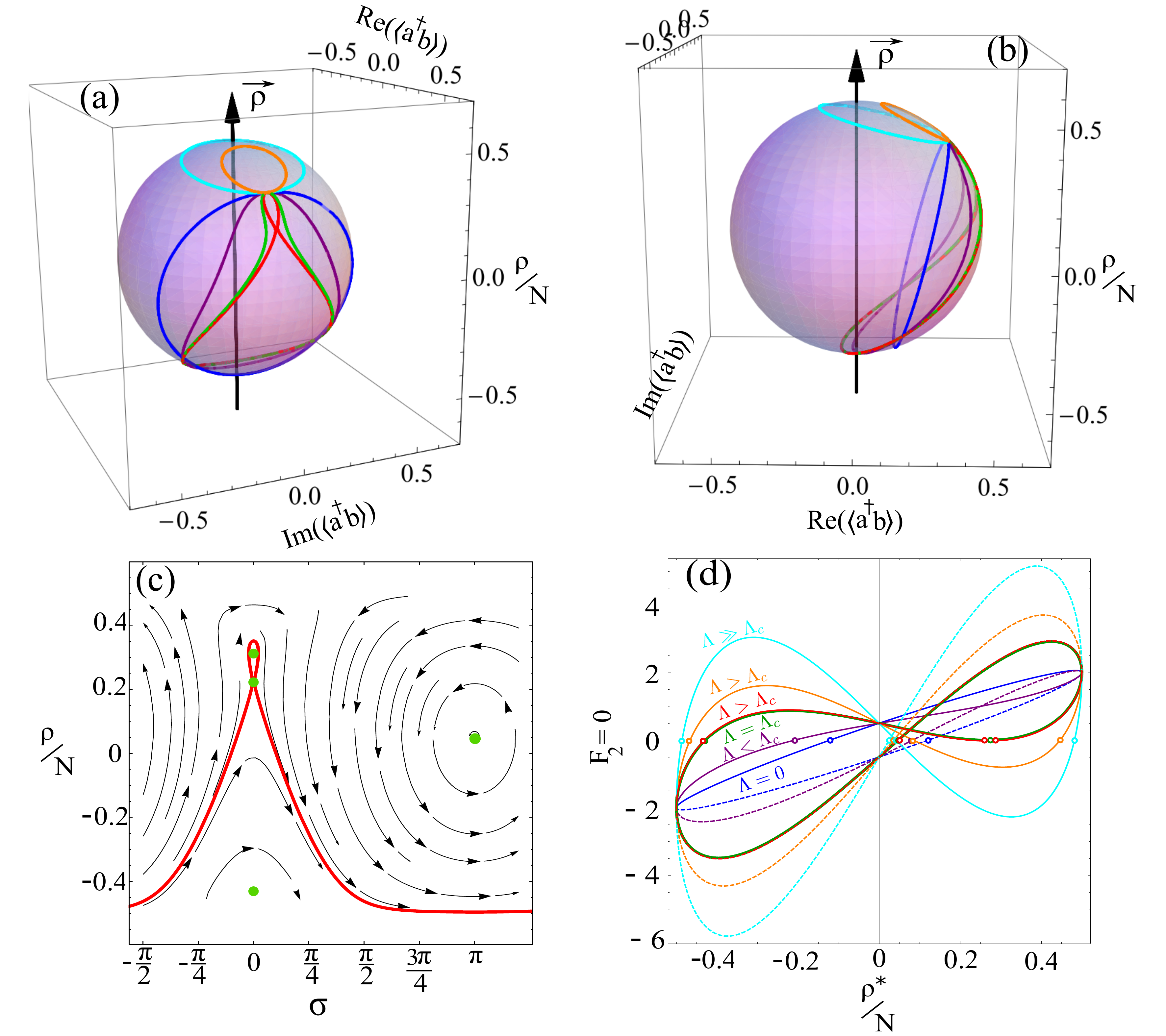}
  \caption{(a) Transition from Rabi to Josephson
    regime.  The Blue circle is the pure Rabi (no interaction)
    regime. Purple is a Rabi-like interacting case,
    with~$0<\Lambda<\Lambda_c$, green is the transition case
    when~$\Lambda=\Lambda_c$, red is a Josephson case with~$\Lambda$
    slightly over $\Lambda_c$, orange and cyan are Josephson
    cases well above $\Lambda_c$.  (b) Same as (a) but as a side view
    of the trajectories to show the cases that do encircle or not the
    $\vec\rho$ axis, corresponding to oscillatory and running relative
    phase, respectively. (c) Phase-space trajectory of the dynamics in
    the Josephson regime with a saddle point at~$\sigma=0$ out of the
    four fixed points. Each~$\Lambda$ yields its own phase-space
    vector field, in which a trajectory is followed depending on the
    initial condition.  (d) Roots of $F_2=0$, that identify the fixed
    point in the population imbalance for the relative phase
    $\sigma=0$ mod~$2\pi$ (solid line) and $\sigma=\pi$ mod~$2\pi$
    (dashed line).  With increasing~$\Lambda$, the number of roots
    changes from two (Rabi regime) to four (Josephson regime).}
  \label{fig:marjul7120100CEST20151207}
\end{figure*}

\begin{figure}[th]
  \centering
  \includegraphics[width=.9\linewidth]{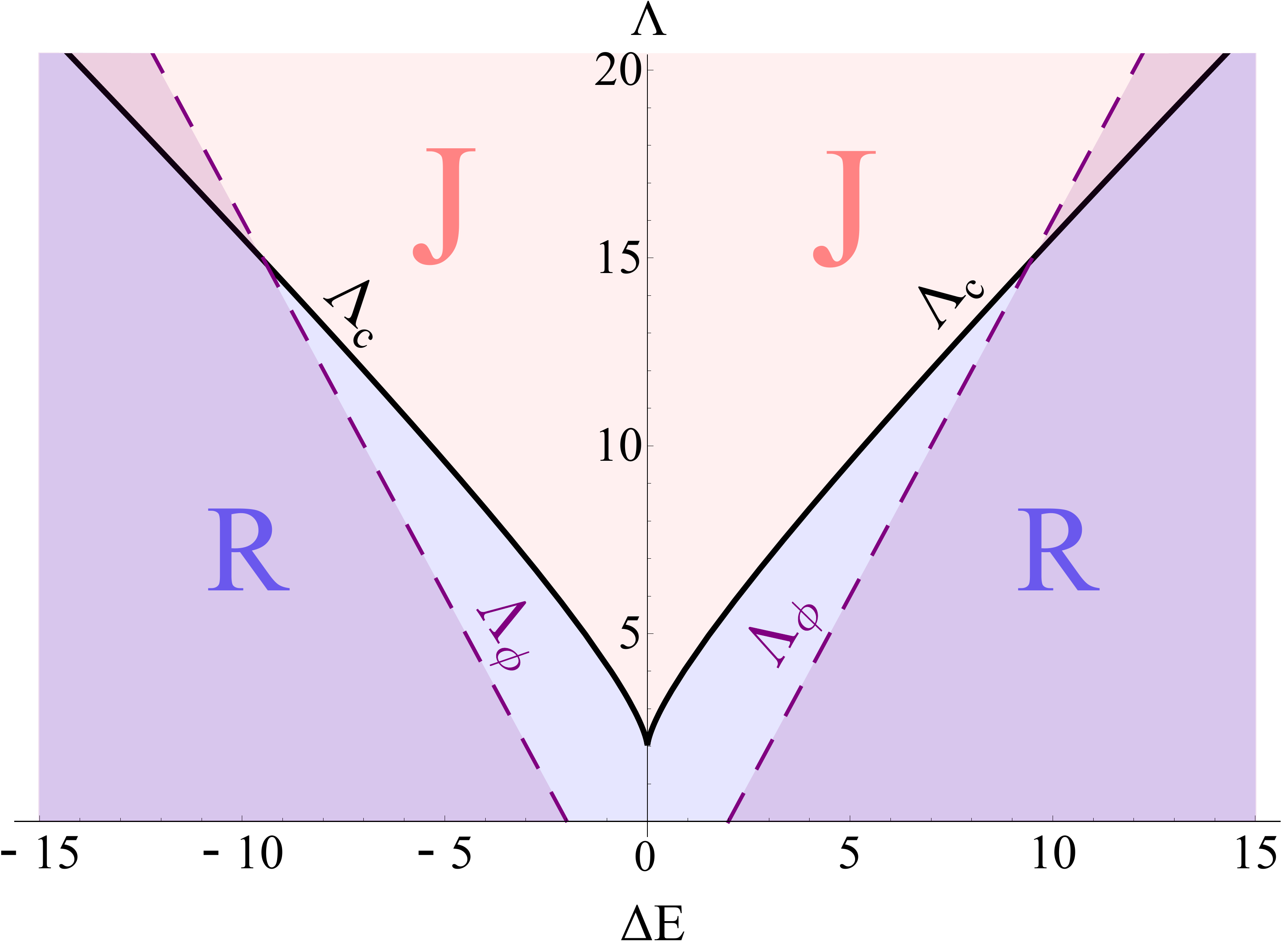}
  \caption{Regions of Rabi (R or blue) and Josephson (J or red) regime
    as a function of $\Lambda$ and $\Delta E$. The frontier,
    $\Lambda_c$ (solid black line) is given by
    Eq.~(\ref{eq:marj282015608}) and provides the general criterion
    for the Josephson regime in presence of detuning. The frontier
    denoted by~$\Lambda_\phi$ (dashed line) separates the
    running-phase regime from the oscillating one (shaded area) for
    the case $(\rho_0=0,\sigma_0=\pi)$ ($\Lambda_\phi$ depends on the
    quantum state and we show here the case of greatest extent for the
    running phase). There is some degree of correlation between
    running phase and Josephson dynamics but neither implies the
    other.}
  \label{fig:marjul7120100CEST93140}
\end{figure}

\subsubsection{Interacting case}

The general interacting case has fixed points solutions in implicit
form:
\begin{equation}
  \label{eq:juefeb11175644CET2016}
  4(\rho^*/N)+e^{i\sigma^*}(\sqrt{1-4(\rho^{*}/N)^2}(\Delta E-2(\rho^*/N)\Lambda))=0\,,
\end{equation}
also for~$\sigma^*\in\{0,\pi\}$. Solutions also exist in closed-form
but are too bulky to give here. The geometrical solution is, in this
case, convenient. It is shown in
Fig.~\ref{fig:marjul7120100CEST20151207}(d) for various values
of~$\Lambda_c$. From the shape of the curve, one can see that there
are two or four fixed points, and this is the criterion one can
unambiguously use to define the Rabi and Josephson regimes,
respectively. This can be quantified by studying the order of the
discriminant of Eq.~(\ref{eq:juefeb11175644CET2016}), yielding the
Josephson regime when it is higher than quadratic in $\rho$.  This
leads us to one important result of this text: the generalized
criterion for Josephson dynamics.  The critical parameter that
separates the Rabi from Josephson regimes in the mean-field
approximation is thus:
\begin{equation}
  \label{eq:marj282015608}
  \Lambda_c=\sqrt{4+\Delta E^2+\frac{6(2\Delta E^2)^{2/3}}{\Xi^{1/3}}+3(2\Delta E^2)^{1/3}\Xi^{1/3}}\,,
\end{equation}
with~$\Xi=4+\Delta E^2+|4-\Delta E^2|$.  In the literature, the
typical configuration reduces to~$\Delta E=0$ and
yields~$\Lambda_c=2$. The diagram in
Fig.~\ref{fig:marjul7120100CEST93140} shows the regions of Rabi (R or
blue) and Josephson (J or red) separated by the $\Lambda_c$ frontier
(black solid line) according to this general criterion. One expects the
Josephson regime to occur with increasing effective interaction
($\Lambda$). However, this is strongly countered by detuning, that
tends to maintain the Rabi regime with a steep increase of the
threshold, that is doubled for a detuning of one time the coupling
strength only. In highly detuned conditions, the Rabi regime
predominates, even with large values of~$\Lambda$.

The fixed points analysis is done for each value of the phase
separately. The $\sigma^{\ast}=0$ solution yields eigenvalues
$\lambda_{\nu}=\pm
i\sqrt{2}{\sqrt{2-\Lambda(1-4\rho^{\ast}\rho^{\ast})^{3/2}}}\big/{\sqrt{1-4\rho^{\ast}\rho^{\ast}}}$
($\nu=1,2$), that imply~$\tau=0$ and, as far
as~$\Lambda\leq\Lambda_c$, $\Delta>0$ meaning that the fixed points
remain centers (these are the $\mathrm{H}_{(\Lambda\leq\Lambda_c)}$
points with nonzero~$\Lambda$ in
Fig.~\ref{fig:marjul7120100CEST83140}). However
for~$\Lambda>\Lambda_c$, one fixed point falls in the region
$\Delta<0$ and becomes a saddle point
($\mathrm{H}_{(\Lambda>\Lambda_c)}$).  For $\sigma^{\ast}=\pi$, the
eigenvalues read $\lambda_{\nu}=\pm
i\sqrt{2}{\sqrt{2+\Lambda(1-4\rho^{\ast}\rho^{\ast})^{3/2}}}\big/{\sqrt{1-4\rho^{\ast}\rho^{\ast}}}$
($\nu=1,2$) which, for all the values of~$\Lambda$, results
in~$\Delta>0$, meaning that all fixed points around~$\sigma^*=\pi$
remain center points ($\mathrm{H}_{(\Lambda\leq\Lambda_c)}$),
regardless of the strength of the interaction. The existence of one
saddle point is thus a criterion to identify the Josephson regime in
presence of detuning. On Fig.~\ref{fig:marjul7120100CEST83140}) is
also superimposed as a shaded area the region of oscillating phase for
the case $\rho_0=0$ and~$\sigma_0=\pi$ (each initial condition yields
its own boundary) separated from the region of running phase by the
dashed purple line~$\Lambda_\phi$. While there is a correlation
between the running phase and the Josephson regime, one neither
implies nor is implied by the other.

Examples of orbits on the Bloch sphere in the Hamiltonian regime are
shown in Fig.~\ref{fig:marjul7120100CEST20151207}(a), starting with
the blue circle that corresponds to the pure Rabi regime
($\Lambda=0$). With increasing interactions, orbits take the shape of
the green trace, that is the frontier between the Rabi and Josephson
regimes. Increasing~$\Lambda$ slightly above the critical value, the
saddle point appears, corresponding to the Josephson regime. The same
orbits are also shown in a side view of the sphere, allowing to see
their enclosing or not of the $\vec\rho$ axis and, correspondingly,
the running or oscillatory-regime of the relative phase.

\section{Out-of-equilibrium (Liouvillian) regime}
\label{sec:marjul7120100CEST201519}

We now consider the out-of-equilibrium dynamics, here in presence of
decay only and in next Section also including pumping.

\subsection{Dynamics}
\label{sec:marjul7120100CEST201520}

\begin{figure}[t]
  \centering
  \includegraphics[width=\linewidth]{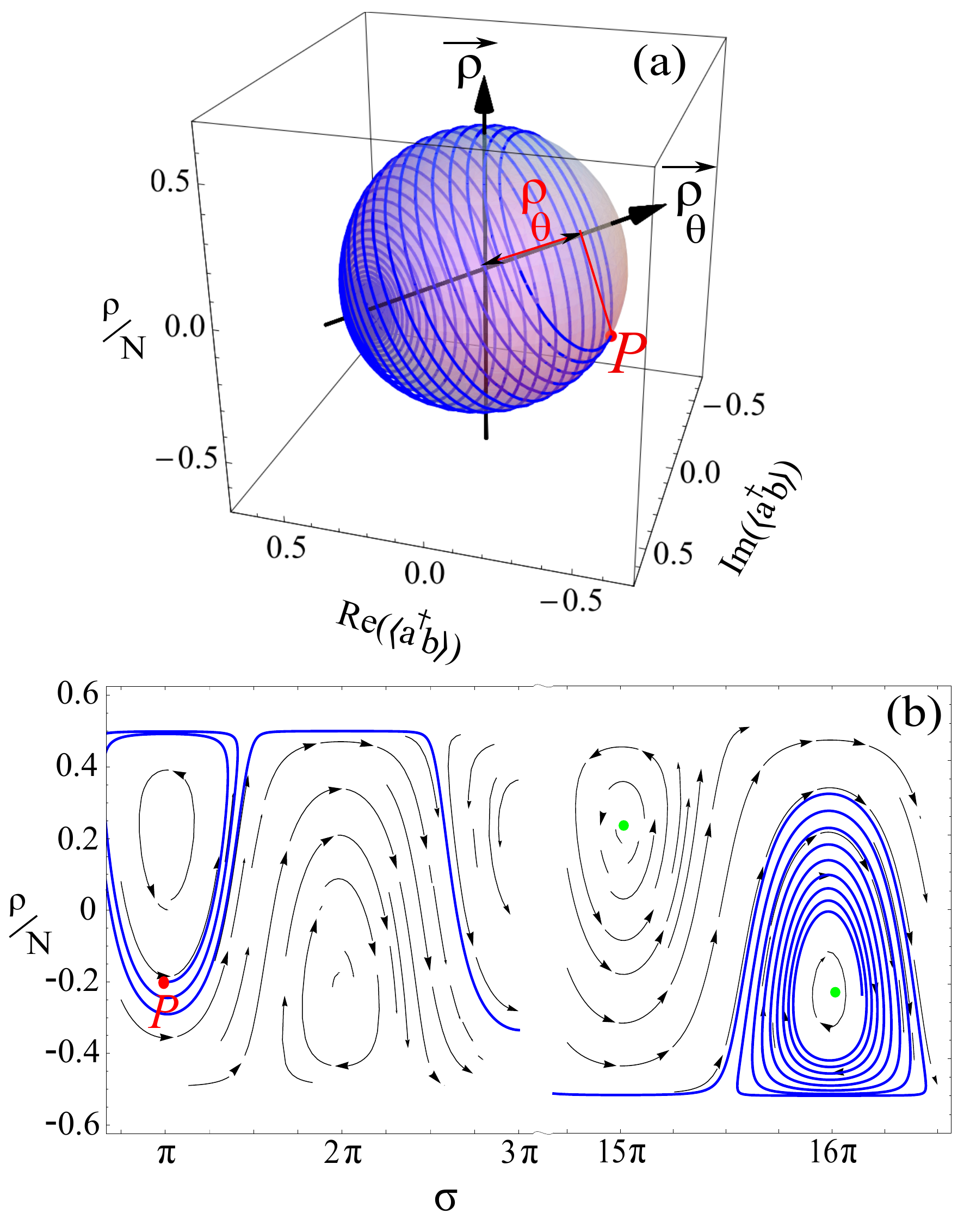}
  \caption{(a) Example of the dynamics in the pure Rabi regime
    ($\Lambda=0$) in a dissipative system. The orbit is an helix on
    the normalized Bloch sphere (Paria sphere), that starts from the
    point~P set by the initial condition and tends toward a steady
    point on the~$\rho_{\theta}$ axes, that remains well defined
    thanks to the normalization despite the steady state being the
    vacuum.  Parameters are the same as in
    Fig.~(\ref{fig:marjul7120100CEST2015123}(h--i). (b) Projection of
    the dynamics on the $(\rho,\sigma)$ space, superimposed on the
    streamlines of the dynamical system.  As compared to the
    Hamiltonian case, the fixed points (displayed as green points) are
    shifted. Starting from~$\sigma_0=\pi$, the spiral gets farther
    from~$\pi$, then drifts as the system enters in the running-phase
    regime, and ultimately gets absorbed by the fixed point
    near~$16\pi$. The left spiral is unstable, while the right one is
    stable.}
  \label{fig:marjul7120100CEST81253}
\end{figure}

We upgrade the Hamiltonian (H) case to include decay by turning to a
Liouvillian (L) description. Considering only decay, this describes
the dynamics of particles with a lifetime, starting from an initial
state, e.g., following a pulsed excitation.  In this regime, as
already commented, one can observe a perplexing switching between the
two regimes of relative phase, shown in
Fig.~\ref{fig:marjul7120100CEST2015123}(h--i). The reason for this
behaviour is readily understood on the normalized Bloch sphere, where
the running or oscillating phase is a topological feature of a
trajectory encircling, or not, the axis of observables.  The
trajectory on this sphere in presence of decay is shown in
Fig.~\ref{fig:marjul7120100CEST81253}~(a). It is helical as it drifts
along the~$\vec\rho_\theta$ axis, from i) the initial point~$P$ which
distance from the center on the $\vec\rho_\theta$ axis is given by
Eq.~(\ref{eq:marene12230820CET2016}) and phase by
\begin{equation}
  \label{eq:marene13230745CET2016}
  \sigma_\theta=\arg[\mathrm{Re}\langle a^{\dagger}b\rangle\frac{\delta}{\sqrt{4+\delta^2}}-\frac{(n_a-n_b)}{4+\delta^2}+i \mathrm{Im}\langle a^{\dagger}b\rangle]\,,
\end{equation}
to ii) one pole of the sphere, still along the $\vec\rho_\theta$ axis,
depending on which particles, $a_\theta$ or $b_\theta$ have the
smaller lifetime.  The distance~$\rho_\theta(t)$ at intermediate times
is easily obtained on physical grounds as:
\begin{equation}
  \label{eq:marene12232914CET2016}
  \rho_\theta(t)=\frac12\frac{n_{a_\theta}(0)e^{-\gamma_{a_\theta}t}-n_{b_\theta}(0)e^{-\gamma_{b_\theta}t}}{n_{a_\theta}(0)e^{-\gamma_{a_\theta}t}+n_{b_\theta}(0)e^{-\gamma_{b_\theta}t}}\,,  
\end{equation}
where $n_{a_\theta}\equiv\langle\ud{a_\theta}a_\theta\rangle$ and
$n_{b_\theta}\equiv\langle\ud{b_\theta}b_\theta\rangle$ follow from
Eqs.~(\ref{eq:marjul7120100CEST20159899}) as
$n_{a_\theta}=n_a\cos^2\theta+n_b\sin^2\theta+2\mathrm{Re}(\langle\ud{a}b\rangle)\cos\theta\sin\theta$,~$n_{b_\theta}=n_a\sin^2\theta+n_b\cos^2\theta-2\mathrm{Re}(\langle\ud{a}b\rangle)\cos\theta\sin\theta$ and
$\gamma_{a_\theta}\equiv\gamma_a\cos^2\theta+\gamma_b\sin^2\theta$,
$\gamma_{b_\theta}\equiv\gamma_a\sin^2\theta+\gamma_b\cos^2\theta$.
Now, in the cases where the $\vec\rho_\theta$ axis is not aligned with
the observable $\vec\rho$ axis---which is all the cases except at
resonance---and if the initial and final points on $\vec\rho_\theta$
are on opposite side of its zero, then the circle will come to
encircle for some time the $\vec\rho$~axis, corresponding to the
running regime of relative phase, until it drifts again on the other
side of the sphere, at which point the system goes to the oscillatory
regime. It can happen that this spiral will pass by the north or south
pole of the $\vec\rho$ axis, which means that in the basis of
observables, one population becomes exactly zero, leading to an
undefined relative phase. This is not, however, compulsory. In other
cases, depending on the interplay between decay and detuning, the
trajectory remains the whole time on one side of the sphere, in which
case the system is always in the oscillating-phase regime and there is
no switching.
 
There are other notable behaviours that are conveniently pictured on
the sphere. At resonance ($\delta=0$) and for dressed states
($\rho_0=\pm N/2$), when~$\Gamma_-=0$, the relative phase starts
at~$\pi/2$ and is then locked at $\pm\pi/2$ forever. This is a
manifestation of optimal strong-coupling with full-amplitude Rabi
oscillations at the Rabi frequency. Moreover, the population imbalance
oscillates in time around~$\rho=0$ while decaying toward zero. Still
at resonance, but now when~$\Gamma_-\neq0$, the relative phase
oscillates in time taking all the values between~$\pm\pi/2$ while the
population imbalance decays faster as compared to the former
case. Out of resonance, $\delta\neq0$, when $\Gamma_-=0$ the relative
phase exhibits the same trend as the Hamiltonian regime, however, the
population decays in time.

\subsection{Classification of fixed points}

The stability analysis in the Liouvillian case shows that the dynamics
is richer and visits extended areas of the stability diagram. This
results in the family of $\mathrm{L}$ points in
Fig.~\ref{fig:marjul7120100CEST83140} that we introduce and discuss
individually below.  This new phenomenology is an important
consideration for polaritons that are inherently finite-lifetime
particles.

The same stability treatment as before but now with the system of
Eqs.~(\ref{eq:marjul7120733CEST201145666}) yields for~$\tau$
and~$\Delta$:
\begin{subequations}
  \begin{align}
    \label{eq:marju21320161843}
    \tau=&-4\Gamma_-\rho^{\ast}/N\,,\\
    \Delta=&(4/(1-4(\rho^*/N)^2))-\Gamma_-^2(1-4(\rho^*/N)^2))\,.\label{eq:marju21320161844}
  \end{align}
\end{subequations}

\begin{figure}[t]
  \centering
  \includegraphics[width=.8\linewidth]{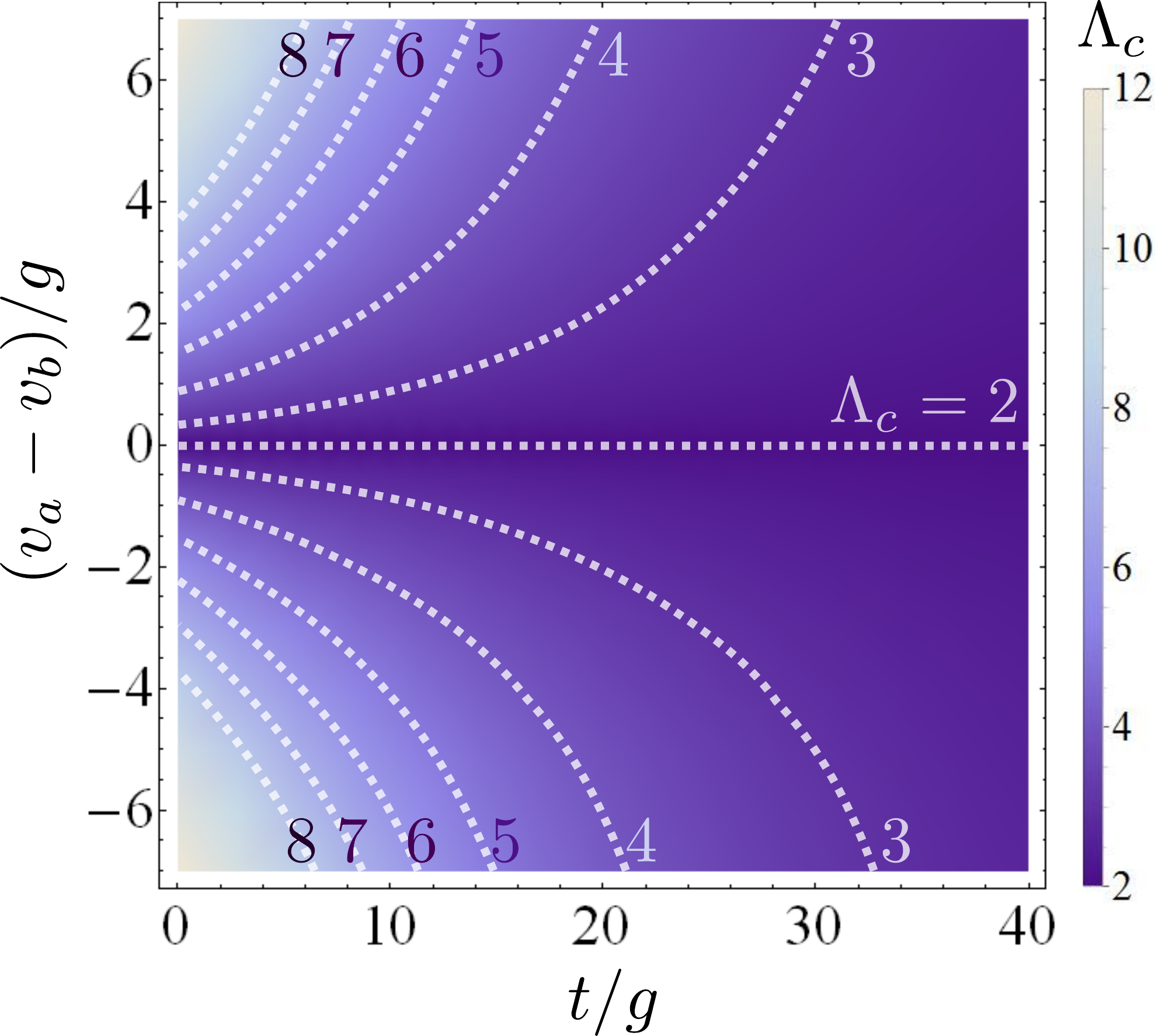}
  \caption{When the interactions of the two condensates are not equal,
    $v_a\neq v_b$, the critical $\Lambda_c$ becomes time-dependent, as
    shown here as a density plot for the case of zero detuning. The
    smallest value is~$\Lambda_c=2$ which is the textbook value for
    the Josephson regime in the case of equal interactions, at
    resonance and without dissipation, as recovered here for this
    particular case. Variations result in an increase of~$\Lambda_c$,
    that decays with time to tend towards this fundamental value.}
  \label{fig:marjul7120100CEST8125312}
\end{figure}

This shows how~$\Gamma_-\equiv\gamma_b-\gamma_a$ (as defined in
Eq.~(\ref{eq:domfeb28122646CET2016}) but here with~$p_a=p_b=0$)
defines new types of fixed points in the $\Delta>0$ region, namely,
the system can also spiral towards its fixed points ($\mathrm{LUD}$
and $\mathrm{LID}_{(\Lambda\leq\Lambda_c)}$ points), as expected from
decay, instead of always orbiting them as before ($\mathrm{LUR}$ and
$\mathrm{LIR}_{(\Lambda\leq\Lambda_c)}$ points).  An example of a
$\mathrm{LUD}$ trajectory, i.e., in the non-interacting (Rabi) regime
and in presence of detuning and decay, is given in
Fig.~\ref{fig:marjul7120100CEST81253}.

From the layout of points in Fig.~\ref{fig:marjul7120100CEST83140},
the stability property of the fixed points in presence of decay thus
remains a good criterion to set apart the Rabi and Josephson regimes.
$\Lambda_c$ is still defined according to
Eq.~(\ref{eq:marj282015608}), but becomes time dependent when $v_a\neq
v_b$ since in this case it depends through $\Delta E$---itself defined
in Eq.~(\ref{eq:domfeb28110505CET2016})---on the total population~$N$
that decays according to $N(t)\approx
N(0)\exp(-(\gamma_a+\gamma_b)t/2)$ (the exact solution is more complex
than this overall pattern; it typically oscillates around this
envelope due to interactions and may exhibit complicated patterns with
abrupt variations in some particular cases, with a dynamics that would
deserve an analysis of its own). The dependence of~$\Lambda_c$ as
function of time and the detuning in interactions, $v_a-v_b$, is shown
in Fig.~\ref{fig:marjul7120100CEST8125312} for the case of bare mode
resonances, $\delta=0$, where it is seen that $v_a=v_b$ makes it
time-independent indeed and pinned to the textbook
value~$\Lambda_c=2$, while an interaction imbalance results in a
dependence of~$\Lambda_c$ similar to that due to detuning
(cf.~Fig.~\ref{fig:marjul7120100CEST93140}). That is, the threshold
for the Josephson regime is increased and decays in time down to the
value~$\Lambda_c$ of Eq.~(\ref{eq:marj282015608}) at long times. The
Rabi regime is therefore always recovered since also~$\Lambda$,
Eq.~(\ref{eq:marmar8144817CET2016}), decays with time, proportionally
to~$N$.  We now discuss the various fixed points in detail, starting
with the non-interacting case ($v_a=v_b=0$) and then considering
interactions. In both cases, however, note that with decay only (no
pumping), the steady state is the vacuum, but is approached in a limit
that is well-defined and that allows the following discussion. Even
more enlighteningly, it can be seen as a fixed point on the normalized
Bloch sphere, showing again the value of this ghost object of varying
radius that unifies the dynamics of relative phase and population
imbalance in a transparent way. To distinguish this variation of
dynamically evolving sphere from the conventional Bloch sphere, we
would propose a dedicated terminology and refer to it as a ``Paria
sphere'' (after the American ghost city).

\subsubsection{Non-interacting case}

In the dissipative Rabi regime, that is, with decay but no
interactions, the fixed points are given by:
\begin{subequations}\label{eq:marju2132016193908}
  \begin{align}
    (\rho^*)^2&=\frac{N^2}{8\Gamma_-^2}[  -4+\Gamma_-^2-\Delta E^2\nonumber\\
    &+\sqrt{\Delta E^4+2\Delta E^2(4+\Gamma_-^2)+(-4+\Gamma_-^2)^2}]\,,\label{eq:marju21320161931}\\
    \sin(\sigma^*)&=\frac12\Gamma_-\sqrt{1-4
      (\rho^*/N)^2}\,.\label{eq:marju21320161932}
  \end{align}
\end{subequations}
Therefore, for zero detuning and $|\Gamma_-|\leq2$, one finds the
fixed points at~$(\rho^*=0,\sigma^*=\sin^{-1}(\Gamma_-/2))$ and even
in the dissipative regime, these fixed points remain centers
($\mathrm{LUR}_{|\Gamma_-|<2}$).  Increasing~$|\Gamma_-|$ makes two
consecutive centers from the set of fixed points approach each other
along the $\rho=0$ axis until they meet when~$|\Gamma_-|=2$ with the
common phase $\sigma=(2k+1)\pi/2$ for integer~$k$, at which point they
become degenerate, as the $\mathrm{LUR}_{(|\Gamma_-|=2)}$ points.
For~$|\Gamma_-|>2$, the fixed points split again but now along the
$\sigma$ axis, as they keep a common value for the phase but depart in
population imbalance according
to~$\rho^*=\pm\sqrt{(-4+\Gamma_-^2)/4\Gamma_-^2}$. Past
$|\Gamma_-|>2$, the fixed points also change their stability property
to become spiral points ($\mathrm{LUR}_{(|\Gamma_-|>2)}$). Beside,
they are now connected by streamlines in the $(\rho,\sigma)$ space,
i.e., starting close from the unstable point brings the system towards
the other point, that is stable.  At non-zero detuning, the fixed
points always are of the spiraling type, $\mathrm{LUD}$. Finally, it
can be shown that the condition $\tau^2-4\Delta<0$ separating spirals
from nodal points is always satisfied, so the system is at most
spiraling.

\begin{figure}[t!]
  \centering
  \includegraphics[width=\linewidth]{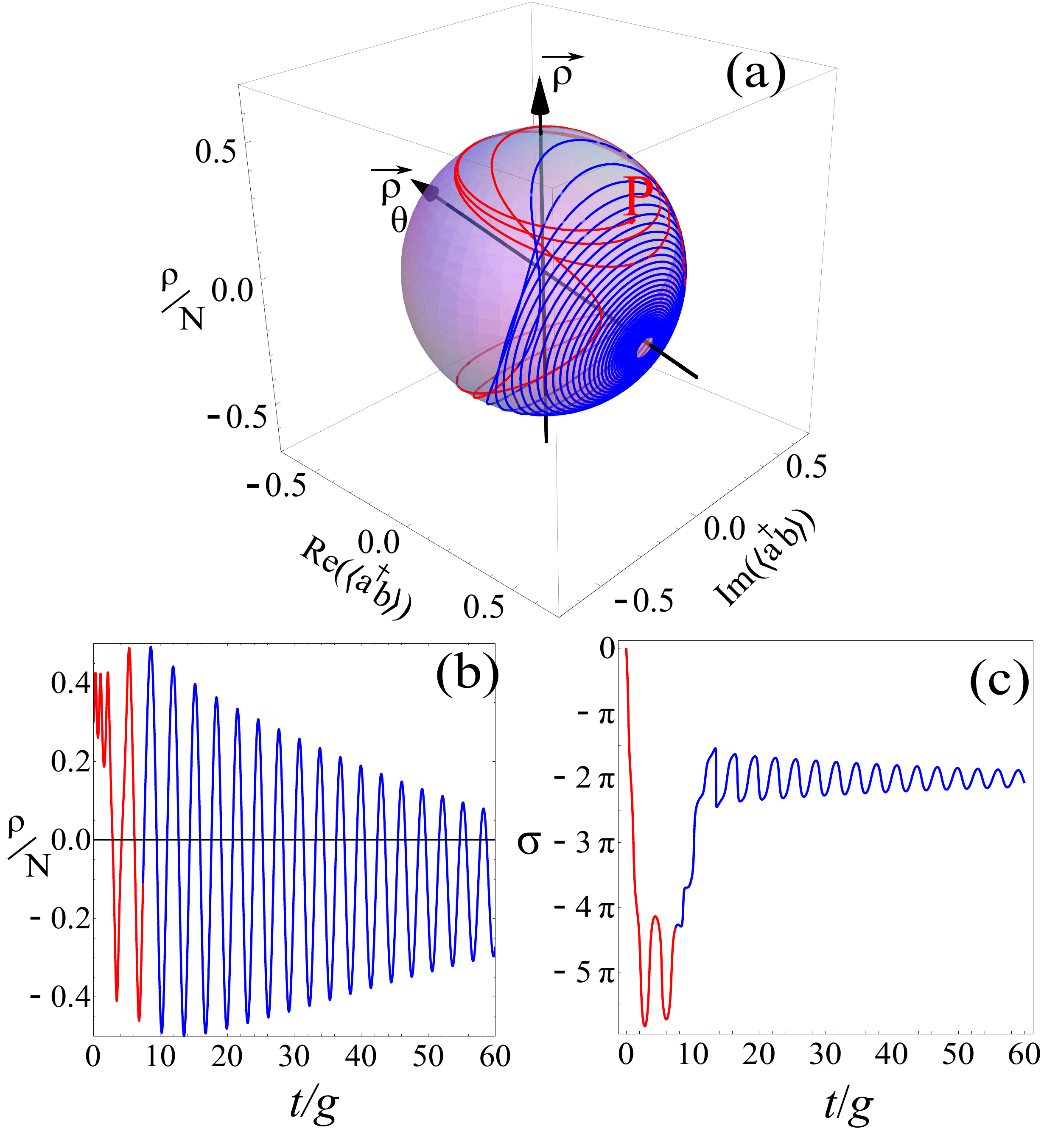}
  \caption{(a) Switching from the Josephson (red trace) to the Rabi
    (blue trace) regimes in an interacting, dissipative system. Here
    we set~$va=vb$. P is the starting point in a system of the
    $\mathrm{LI}_{(\Lambda>\Lambda_c)}$ type. Due to decay, the system
    eventually switches to the Rabi regime, with the dynamics ending
    at a point on the~$\rho_{\theta}$~axes. (b) The population
    imbalance shows two kinds of self-trapping, one at early times (in
    red) that is induced by the interactions, the other at later times
    (in blue) that is induced by detuning. (c) The relative phase also
    exhibit a switching between the oscillating- and running-phases,
    in a way such that all the four possible combinations
    (Josephson--Running-phase; Josephson--Oscillating-phase;
    Rabi--Running-phase and Rabi--Oscillating-phase) happen in
    succession. Parameters: $\rho_0=0.3N$, $\sigma_0=0$, $\Delta
    E=0.5$, $\Lambda(0)=12$, $\gamma_a=0.25g$ and $\gamma_b=0.05g$. }
  \label{fig:marjul7120100CEST81257}
\end{figure}

\subsubsection{Interacting case}

Clearly, with decay, the total number of particles decays with time,
and even if starting in the Josephson regime, ultimately the system
gets into the Rabi regime where tunneling (or coupling) dominates over
interactions.  That is, the system eventually follows the dynamics of
Eqs.~(\ref{eq:marjul7120733CEST201145666}), that yields the steady
state of the previous Section through
Eqs.~(\ref{eq:marju2132016193908}).  Such a transition between the two
regimes might in fact be the clearest evidence of the Josephson regime
in a dissipative context.  This is illustrated with the example of the
dynamics in Fig.~\ref{fig:marjul7120100CEST81257}, that starts from a
point in the Josephson regime, i.e., a
$\mathrm{LI}_{(\Lambda>\Lambda_c)}$ point in
Fig.~\ref{fig:marjul7120100CEST83140}. Then $\Lambda$ decays along
with the number of particles as time passes, and the helix drifts till
$\Lambda=\Lambda_c$ at which point the dynamics switches to the Rabi
regime (now plotted in Blue as compared to Red in the Josephson
regime), and subsequently spirals along the~$\rho_{\theta}$ axis.
Such a switching gives rise to two kinds of ``population trapping'',
i.e., nonzero time-averaged population imbalance
$\langle\rho\rangle$. One trapping is caused by the interactions, and
occurs in the Josephson regime, while another type of trapping is
caused instead merely by detuning, and occurs in the Rabi regime. Just
as the distinction between the Rabi and the Josephson regimes might be
arduous to make in cases where interactions, detuning and decay
compete, also the type of trapping could be ambiguous. A decay-induced
switching of regime is shown in
Fig.~\ref{fig:marjul7120100CEST81257}(b), with two types of trapping
on both sides of the
switching. Figure~\ref{fig:marjul7120100CEST81257}(c) shows the
behavior of the relative phase versus time, changing from
running-phase to oscillatory as $\Lambda$ decays, up to the switching
time, after which point the phase runs again but in the Rabi regime
until, eventually, it is brought back to the oscillating-phase regime
(still in the Rabi regime).  With this example, one can see the
diversity of the possible regimes, both for the dynamics of the
relative phase (running and oscillatory) and for the type of
the oscillations (Josephson and Rabi).  While the interaction mediates the
change of regime, decay mediates the change in the phase dynamics. In
total, we have four combinations that succeed to each others, that
illustrate well the complexity of the phenomenon when considered in
its full generality.

\begin{figure*}[t]
  \centering
  \includegraphics[width=\linewidth]{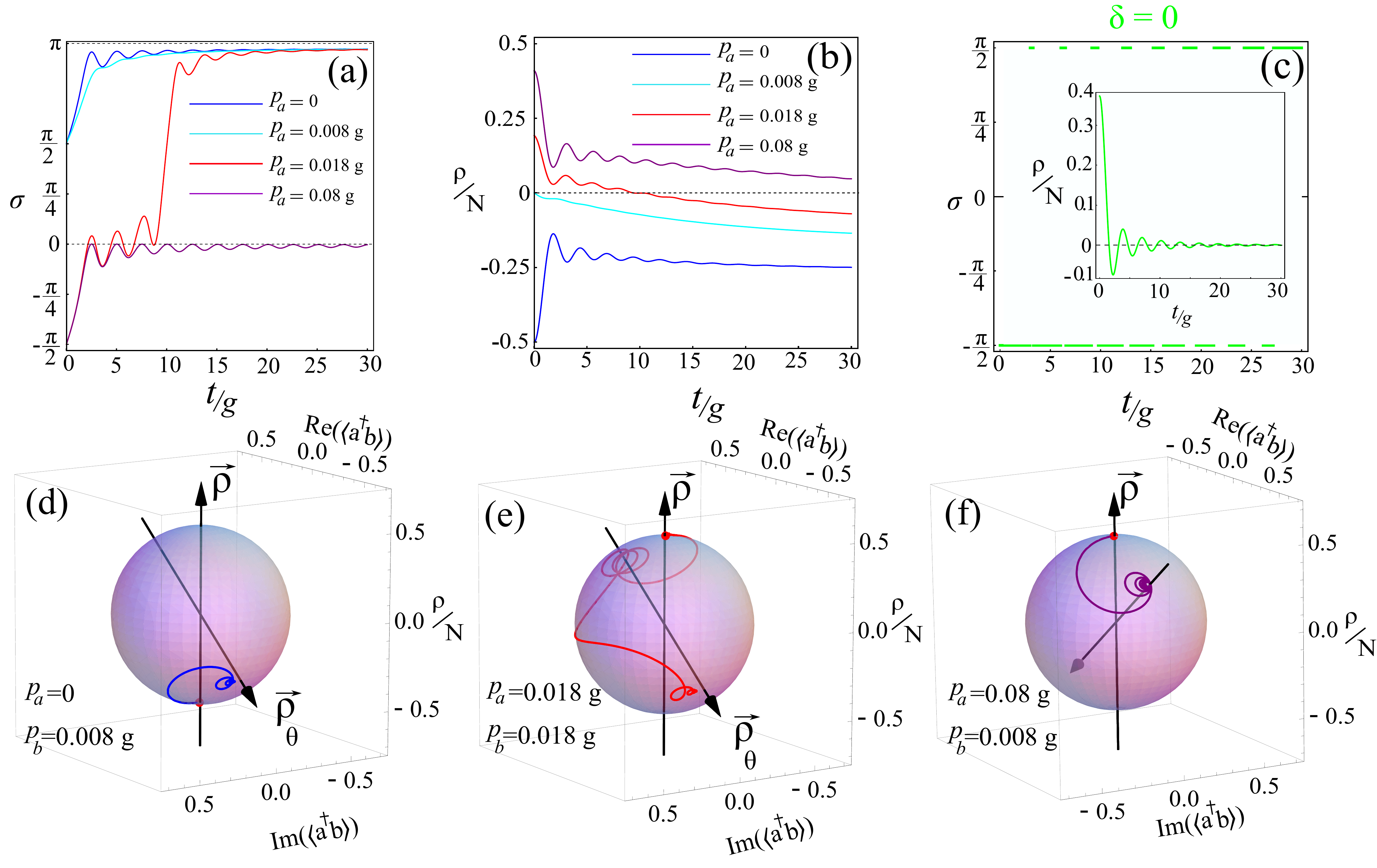}
  \caption{(a) Relative phase and (b)~population imbalance in a
    dissipative system with pumping but in absence of
    interactions. Parameters: $\delta=1.5$ and
    $p_b=8\times10^{-3}g$. The relative phase only display the
    oscillatory regime even though it can jump between fixed
    points. (c) Relative phase at resonance ($\delta=0$), in which
    case, for any pumping ratio, it is a two-valued sawtooth
    function. In this case, the population imbalance goes toward
    zero. (d--f) Dynamics on the Bloch sphere for the three different
    pumping ratio considered in panel~(a). The trajectories never
    encircle the vertical~$\vec\rho$ axis, thereby suppressing the
    running-regime of the phase. Parameters for all panels:
    $\gamma_a=0.2g$ and $\gamma_b=0.02g$.}
  \label{fig:marjul7120100CEST20159}
\end{figure*}

\section{Pumping}
\label{sec:marjul7120100CEST201521}
\subsection{Dynamics}

We now succinctly consider the dynamics of relative phase from
Eqs.~\ref{eq:marjul7120100CEST20159876} with nonzero~$p_a$ and~$p_b$,
that is, in presence of incoherent pumping. The general case would
bring us too far astray and we therefore limit our discussion to the
pure Rabi regime (non-interacting case). We first consider the
transient, starting from the vacuum and then briefly describe the
steady state situation before giving it full attention in next
section.

Starting from vacuum, the relative phase is ill-defined at~$t=0$. With
pumping, populations in both states increase and a relative phase is
established. Note that while~$S=0$ at all times since both~$\langle
a\rangle$ and~$\langle b\rangle$ remain zero under incoherent
pumping---that randomizes the phase of each mode---the
cross-correlator~$\sigma=\langle\ud{a}b\rangle$ is well defined and
gets interconnected to particles tunneling in a way similar to the
Hamiltonian case.  This supports the idea that there is no absolute
phase for the condensate but a degree of cross-coherence, or
correlation, which, for the sake of convenience, we can still refer to
as a relative phase (its argument, $\sigma$ is for all purposes a
relative phase). This phase sets itself at the value~$\pm\pi/2$ at
early times depending on the ratio~$p_a/p_b$.
Figure~\ref{fig:marjul7120100CEST20159}(a-b) shows its subsequent
evolution along with the population imbalance. When $p_a/p_b\leq 1$,
the relative phase starts from~$\approx\pi/2$ and then evolves
oscillatory in time towards its steady value
of~$\approx\pi$. Meanwhile, the population imbalance remains negative
while its absolute value increases. When~$p_a/p_b>1$, the relative
phase starts from~$\approx-\pi/2$. Also, its time evolution is
different than the case~$p_a/p_b\leq 1$. As shown in
Fig.~\ref{fig:marjul7120100CEST20159}(a-b) for~$p_a=0.018g$, there are
oscillations around zero followed by a jump to oscillations
around~$\pi$. This behavior is connected to the population imbalance
that changes sign (from positive to negative), crossing zero.  For the
oscillations in the relative phase to end up with its steady state
around~$0$, the population imbalance must remain positive. At
resonance ($\delta=0$), the relative phase remains~$\pm\pi/2$ for any
pumping ratio, and the population imbalance shows damped oscillations
around zero. This is shown in
Fig.~\ref{fig:marjul7120100CEST20159}(c).

Panels (d-f) of Fig.~\ref{fig:marjul7120100CEST20159} shows the
corresponding dynamics on the Bloch sphere for three pumping
ratios. The trajectory in each case starts from the point close to the
south or north pole of the observable axis, then drifts toward a
steady point near the eigenstate axis. As can be seen in the figures,
the trajectory is immediately brought from the observable axis to the
eigenstate one, and thus has no chance to loop around or intersect the
$\vec\rho$ axis.  As a consequence, the relative phase only displays
damped oscillations.  Consequently, the running-phase regime is
suppressed by incoherent pumping. Finally, we have considered
non-interacting systems only but also a fairly simple model of
pumping, through Lindblad operators that are the direct counterpart of
spontaneous decay. Voronova \emph{et al.} have reported interesting
instabilities akin to Kapitza's pendulum in driven interacting
Rabi--Josephson systems with a more elaborate model of pumping
(through a reservoir)~\cite{arXiv_voronova16a}, hinting the complexity
of the more general cases.

\subsection{Classification of fixed points}

The stability analysis in presence of pumping brings some qualitative
novelties due to the randomization of the phase.  First, the fixed
points now lie in a four-dimensional space instead of two before,
since in the steady state the phase acquires a definite complex value,
adding two dimensions and making obsolete the criterion of running vs
oscillating phase on the Paria (normalized Bloch) sphere.  Solving
Eqs.~(\ref{eq:marjul7120100CEST20159876}) in the steady state
(with~$v_a=v_b=0$), one finds:
\begin{subequations}
  \label{eq:marju21320161939082016312}
  \begin{align}
      &n^*=
      \begin{aligned}[t]
        &[p_a(-4\Gamma_++(\Gamma_--\Gamma_+)(\delta^2+\Gamma_+^2))\\
        &-p_b(4\Gamma_++(\Gamma_-+\Gamma_+)(\delta^2+\Gamma_+^2))]/Y\,, \label{eq:marju213201619312001631213}
      \end{aligned}
\\
    &\rho^*=[(\delta^2+\Gamma_+^2)(\Gamma_+(p_b-p_a)+\Gamma_-(p_a+p_b)]/2Y\,, \label{eq:marju213201619312001631212}\\
    &\mathrm{Re}[\langle a^{\dagger}b\rangle^*]=-\delta(\Gamma_+(p_b-p_a)+\Gamma_-(p_a+p_b))/Y\,,
  \label{eq:marju2132016193120016312}\\
  &\mathrm{Im}[\langle a^{\dagger}b\rangle^*]=-(\Gamma_+/\delta)\mathrm{Re}[\langle a^{\dagger}b\rangle^*]
  \,,\label{eq:marju213201619322016312}
\end{align}
\end{subequations}
where~$Y$~stands
for~$(\Gamma_-^2-\Gamma_+^2)(\delta^2+\Gamma_+^2)-4\Gamma_+^2$.  Note
that
Eqs.~(\ref{eq:marju2132016193120016312}--\ref{eq:marju213201619322016312})
are more simply expressed as $\sigma^*=\arctan(-\Gamma_+/\delta)$. 
While the dimension of the space is larger, there is however a single
fixed point, due to the unicity of the steady state solution. The
stability of this point follows from the eigenvalues of the Jacobian
matrix, these being:
\begin{equation}
  \label{eq:marmar15183832CET2016}
  \lambda_{pq}=\Gamma_++pi\sqrt{qX+\sqrt{X^2+4\Gamma_-^2\delta^2}}\,,
\end{equation}
where $X\equiv \Gamma_-^2-\delta^2-4$ and where~$p$ and~$q$ take the
values $\pm1$ (we label them with the sign only, so that, e.g.,
$\lambda_{+-}=\Gamma_++i\sqrt{-X+\sqrt{X^2+4\Gamma_-^2\delta^2}}$).
The corresponding eigenvectors $v_{pq}$ provide the directions in the
$(\rho, n, \sigma)$ four-dimensional space along which the system
flows when slightly perturbed. There is no obvious geometrical
features to characterize $v_{pq}$. The stability properties are the
following: if $\Gamma_+<0$ and $Y<0$, the fixed point is stable. If
only~$\Gamma_+<0$ is satisfied, the subspace spanned by~$\left\lbrace
  v_{--},v_{-+},v_{+-}\right\rbrace $ is stable while any combination
involving $v_{++}$ is unstable. On the contrary, if only~$Y<0$ is
satisfied, the only stable subspace is spanned by~$v_{++}$ while any
other possible linear superposition yields an unstable
point. For~$\Gamma_+>0$, the dynamics is generally unstable, and,
interestingly, can feature saddle-type of instability, that in the
Hamiltonian or dissipative regime (without pumping), was used as a
criterion for the Josephson regime, where the dynamics is ruled by the
(weak) interactions. Here the system has no interactions, but can
still manifest this type of saddle instability and in different ways,
for instance when the condition $\Gamma_+>0$ and $Y<0$ is met.  The
presence of a saddle-type of instability in a non-interacting system
may be disconcerting, because this served in the previous cases as a
robust criterion to identify the Josephson regime, so arguably doubts
may arise on what precisely defines the Josephson regime in the most
general situation.  One could then look for a deeper characterization
to establish such a general criterion when also including pumping, for
instance, through the number of fixed points, or one could also
upgrade the Josephson regime to the realm of pumping non-interacting
systems. However, these various approaches, although they match the
facts, lack a clear physical motivation, so we leave it an open
question whether a general definition is suitable in the most general
case that combines pumping, decay, detuning and interactions.  Another
case worthy of interest is $\Gamma_+=0$ and $Y=0$, that results in
pure imaginary eigenvalues. This gives rise to a center subspace, that
results in a type of bifurcation known as a ``transcritical
bifurcation''~\cite{strogatz_book94a}, that is, the fixed point
exchanges its stability when passing by~$\Gamma_+=0$ or~$Y=0$.
 
\begin{figure}[tbp]
\centering
\includegraphics[width=\linewidth]{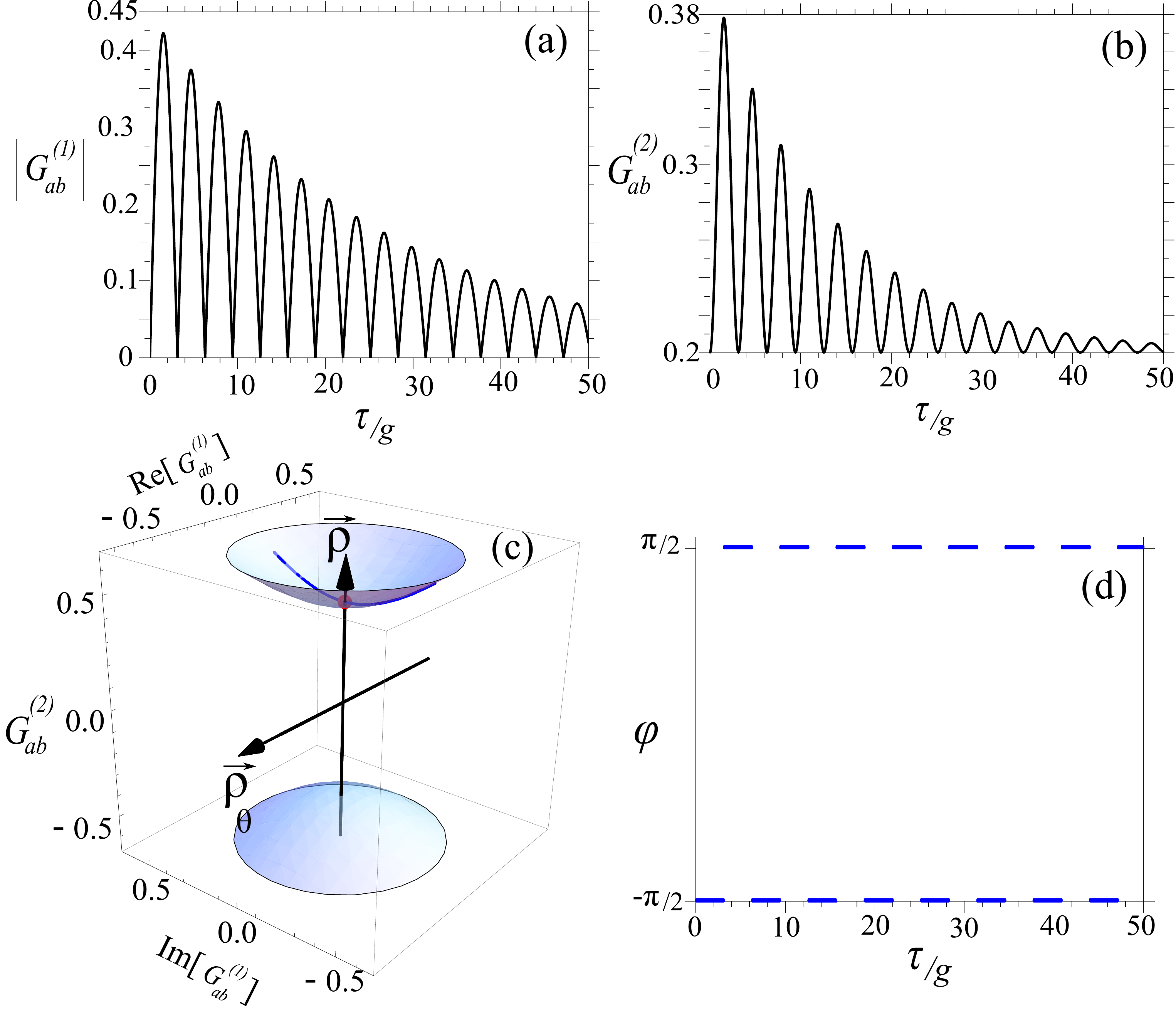}
\caption{\label{fig:marjul7120100CEST201510} (a--b) $\tau$--Dynamics
  of the two-time correlators $G^{(1)}_{ab}$ and $G^{(2)}_{ab}$ in the
  steady state of a dissipative system at resonance and without
  interactions. (c) In autocorrelation time, rather than a sphere, the
  trajectory lies on the upper sheet of an hyperboloid. A red point
  indicates the starting point.  (d) Phase of $G^{(1)}_{ab}$, in the
  oscillating-regime.}
\end{figure}
\begin{figure}[tbp]
\centering
\includegraphics[width=\linewidth]{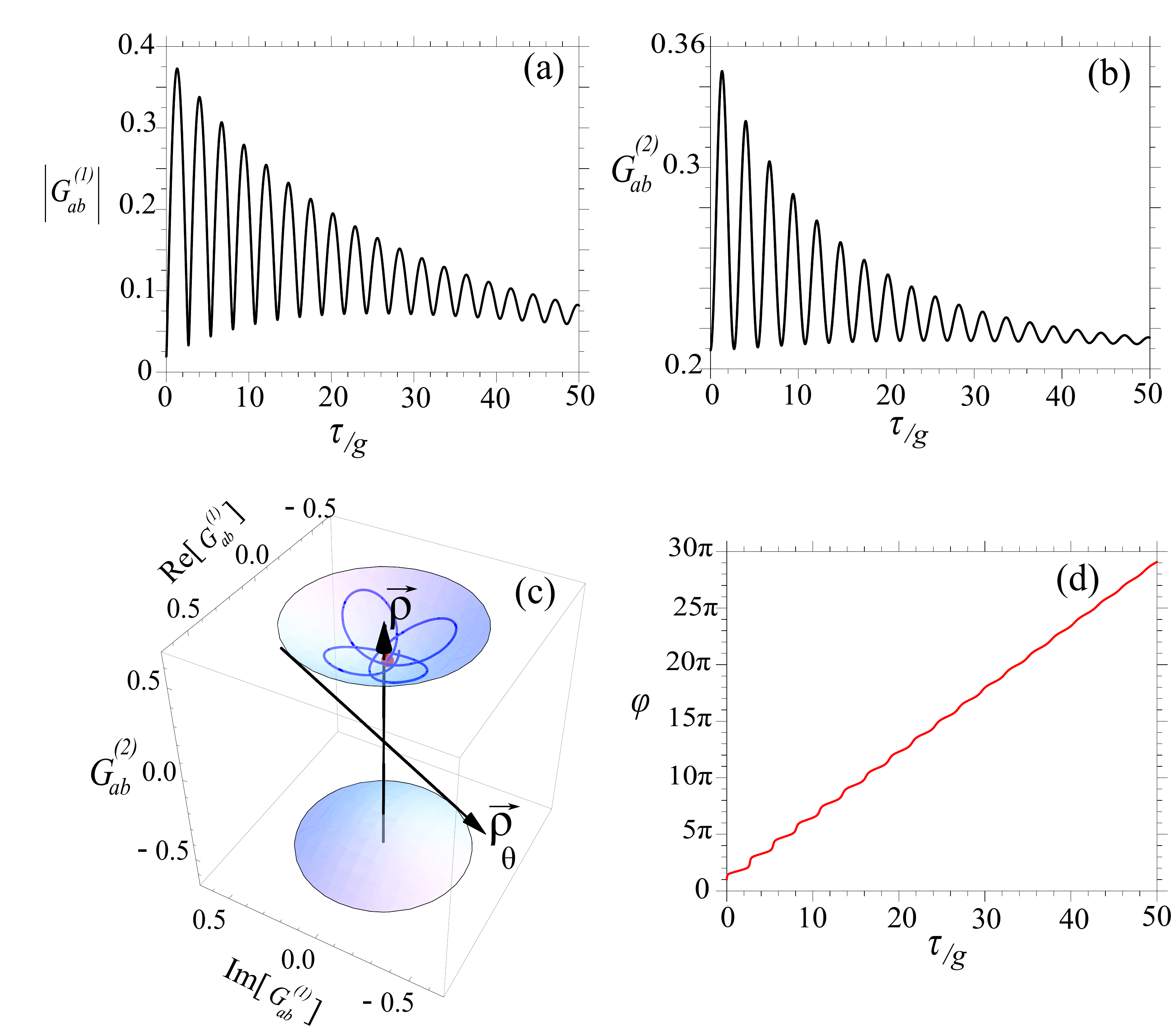}
\caption{\label{fig:marjul7120100CEST201539} Courresponding
  $\tau$--Dynamics of $G^{(1)}_{ab}$ and $G^{(2)}_{ab}$ of
  Fig.~\ref{fig:marjul7120100CEST201510} but in a detuned system.  (c)
  The dynamics on the hyperboloid is now open and as a result, (d),
  the the phase of $G^{(1)}_{ab}$ is in the running-regime. }
\end{figure} 

\subsection{Dynamics in autocorrelations}
\label{sec:marjul7120100CEST201524}

We have clarified the dynamics in real time thanks to a normalized
Bloch sphere, and articulated the fundamental links between the phase
and population imbalance in both pure Hamiltonian dynamics and in the
transient dynamics of a driven dissipative system.  In this section,
we explore possible similar relationships in a steady state situation,
where the~$t$ dynamics has converged to constant values by definition,
with only dynamics in the correlations remaining.  The natural
operators to consider are:
\begin{subequations}
  \label{eq:marjul7120100CEST201525}
  \begin{align}
    G^{(1)}_{ab}(\tau)&=\lim_{t\rightarrow\infty}\langle a^{\dagger}(t+\tau)b(\tau)\rangle\label{eq:marjul7120100CEST201526}\,,\\
    G^{(2)}_{ab}(\tau)&=\lim_{t\rightarrow\infty}\langle a^{\dagger}(t)b^{\dagger}(\tau+t)b(\tau+t)a(t)\rangle\,.\label{eq:marjul7120100CEST201527}
  \end{align}
\end{subequations}
The former, the crossed first order correlation
function~$G^{1}_{ab}(\tau)$, is related to coherence between the
states and suggests as an extension for steady states of~$\sigma$, the
relative phase in autocorrelation time
$\varphi\equiv\arg(G^{(1)}_{ab})$. The latter is related to
fluctuations of the population imbalance in autocorrelation time.
Correspondingly, the question pauses itself whether these two
observables are geometrically connected or not.  Using
Eqs.~(\ref{eq:marjul7120100CEST20159876}) and the quantum regression
theorem, the following relationship can be obtained:
\begin{multline}\label{eq:marjul7120100CEST201533}
  G^{(1)}_{ab}(\tau)=\lim_{t\rightarrow\infty}\frac{1}{\omega}\exp((\delta+i \Gamma_+)\tau /2)\\\big[\langle a^{\dagger}(t)b(t)\rangle(\omega\cos(\omega\tau/2)+(i\delta+\Gamma_-)\sin(\omega\tau/2))\\-2i\langle a^{\dagger}(t)a(t)\rangle\sin(\omega\tau/2)\big]\,,
\end{multline}
where we have introduced a complex frequency
$\omega=\sqrt{4+(\delta-i\Gamma_-)^2}$. By also writing the equation
of motion for~$G^{(2)}_{ab}(\tau)$, it can be shown that
Eqs.~(\ref{eq:marjul7120100CEST201525}) are related through the
relation:
\begin{equation}
  \label{eq:marjul7120100CEST201534}
  G^{(2)}_{ab}(\tau)=|G_{ab}^{(1)}(\tau)|^2+\lim_{t\rightarrow\infty}\frac{N^2}{4}\,.
\end{equation}
This shows that $G^{(1)}_{ab}$ and $G^{(2)}_{ab}$ are coupled indeed,
with a possible similar Josephson interpretation that one is driving
the other. However, their connection is through a two-sheet
hyperboloid and is thus completely different than the the dynamics of
real-time observables, even for the transient dynamics, where
variables are connected via a sphere of variable radius (the Paria
sphere).

Figure~\ref{fig:marjul7120100CEST201510} shows the trajectory in the
hyperboloid at resonance ($\delta=0$). In this case, the relative
phase is a two-valued function, oscillating between~$\pm\pi/2$
(Fig.~\ref{fig:marjul7120100CEST201510}~(d)), which corresponds to the
the oscillatory regime of the relative phase.  Correspondingly,
$G^{(1)}_{ab}$ and $G^{(2)}_{ab}$ oscillate in time with a decay
toward zero for~$G^{(1)}_{ab}$ and $N^2/4$ for
$G^{(2)}_{ab}$. Comparing with panel~(d), it is observed that
$\varphi$ changes value whenever $G^{(1)}_{ab}$ becomes zero.  This
shows a similar behaviour than in the real-time behaviour where
whenever the relative phase becomes ill-defined due to one state
becomings zero avoid, it changes its regime. Here, ill-defined
coherence changes the value of phase instead. Simultaneously,
$G^{(2)}_{ab}$ reaches~$N^2/4$, which is the point of lowest possible
fluctuations. The corresponding trajectory on the hyperboloid in
panel~(c) shows a simple line (in the curved space)
with~$\mathrm{Re}[G^{(1)}_{ab}]=0$ and~$G^{(2)}_{ab}$ swinging around
the lowest point of the hyperboloid.  The detuned case ($\delta\neq0$)
is shown in Fig.~\ref{fig:marjul7120100CEST201539}. This time, as seen
in panel~(d), the relative phase is running. Both $G^{(1)}_{ab}$ and
$G^{(2)}_{ab}$ oscillate and decay toward different steady points as
compared to resonance. The trajectory on the hyperboloid shows this
time an open orbit, encircling the hyperboloid $G^{(2)}_{ab}$ axis
without ever touching it.  In contrast to the real-time dynamics on
the Paria sphere, in autocorrelation time, there is a reduced
phenomenology and, in particular, no preferred change of basis, e.g.,
there is no counterpart of a regime of drifting phase out of resonance
and two-valued $\pm\pi/2$ phase at resonance.  This is due to the
Hamiltonian dynamics being washed out by the incoherent pumping.

\section{Conclusions}

In conclusion, we have generalized and unified the problem of Rabi and
Josephson oscillations between two weakly interacting condensates to
include i) detuning, ii) different interactions for each condensate
and iii) decay. We have also overviewed the situation with pumping,
which, however, requires a dedicated analysis of its own.  Our results
show that even at the simplest mean-field level, such a fundamental
problem had kept some fundamental features hidden through the
particular cases that had been focused on so far.  For instance, the
relative phase $\sigma$ and population imbalance $\rho$ appear in the
most general case as two sides of the same coin without one driving
the other, and their qualitative behaviour depends on a choice of
representation, with a basis that can always be chosen in which there
is a linear phase drift in the pure Rabi regime.  At such, the
behaviour of the relative phase can not be associated to neither the
Josephson nor the Rabi regime exclusively, although it bears some
degree of correlation with it, and is instead a topological
feature. Similar restrains apply to self-trapping.  Such relationships
are elegantly captured on a Bloch sphere of varying radius (that we
termed Paria sphere) that clarifies an otherwise perplexing dynamics
such as a shift of the regime of relative-phase from oscillating to
running and oscillating again.  An unambiguous general criterion to
identify the Rabi ($\Lambda<\Lambda_c$) and Josephson
($\Lambda>\Lambda_c$) regimes has been provided through the critical
effective interaction~$\Lambda_c$, Eq.~(\ref{eq:marj282015608}), that
generalizes the case found in the literature at resonance and for
equal interactions, in which case~$\Lambda_c=2$.  In the Hamiltonian
case, when~$\Lambda<\Lambda_c$, there are two fixed points that are
centers for the dynamics.  When~$\Lambda>\Lambda_c$, there four fixed
points, the one at~$\sigma=0$ and lying between the two other fixed
points being a saddle point, with all other points being
centers. Similar analyses have been undertaken in the Liouvillian case
and are summarized in Fig.~\ref{fig:marjul7120100CEST83140}.  Since in
this case the number of particles dies in time, the system always
eventually shifts to the Rabi regime. In the case of different
interaction strengths, also the critical $\Lambda_c$ becomes time
dependent. Rather than the observation of mere oscillations and/or
population trapping, a clear identification of the Josephson regime in
finite-lifetime particles can instead be made by observing the
transition in time from one regime to the other.

\begin{acknowledgments}
  We thank N.~Voronova for discussions.  Funding by
  the ERC POLAFLOW project No. 308136 is acknowledged.
\end{acknowledgments}

\bibliographystyle{naturemag}
\bibliography{ref4prb,Sci,books,arXiv}

\end{document}